\documentclass[sigconf]{acmart}

\renewcommand\footnotetextcopyrightpermission[1]{}

\usepackage{amsmath}    
\usepackage{amssymb}
\usepackage{wasysym}
\usepackage{geometry}
\usepackage{mathrsfs}
\usepackage{graphicx}
\usepackage{subfigure}
\usepackage{enumerate}
\usepackage{bbm}
\usepackage{amsthm}
\usepackage{setspace}
\usepackage{flushend}
\usepackage{stackrel}
\usepackage{caption}
\usepackage{url}
\usepackage{hyperref}

\newcommand{\ud}{\mathrm{d}}
\allowdisplaybreaks[1]

\begin{document}
\title{Loyalty Programs in the Sharing Economy: \\ Optimality and Competition}

\author{Zhixuan Fang}
\affiliation{%
	\institution{Tsinghua University}
	\city{Beijing}
	\postcode{100084}
	\country{China}
}
\email{fzx13@mails.tsinghua.edu.cn}

\author{Longbo Huang}
\affiliation{%
	\institution{Tsinghua University}
	\city{Beijing}
	\postcode{100084}
	\country{China}
}
\email{longbohuang@tsinghua.edu.cn}

\author{Adam Wierman}
\affiliation{%
	\institution{California Institute of Technology}
	\city{Pasadena}
	\state{CA}
	\postcode{91125}
	\country{USA}
}
\email{adamw@caltech.edu}

\begin{abstract}
Loyalty programs are  important tools for sharing platforms seeking to grow supply.
Online sharing platforms use loyalty programs to heavily subsidize resource providers,  encouraging participation and boosting supply.
As the sharing economy has evolved and competition has increased, the design of loyalty programs has begun to play a crucial role in the pursuit of maximal revenue.
In this paper, we first characterize the optimal loyalty program for a platform with homogeneous users. We then show that optimal revenue in a heterogeneous market can be achieved by a class of multi-threshold loyalty program (MTLP) which admits a simple implementation-friendly structure. 
We also study the performance of loyalty programs in a setting with two competing sharing platforms, showing that the degree of heterogeneity is a crucial factor for both loyalty programs and pricing strategies. Our results show that sophisticated loyalty programs that reward suppliers via stepwise linear functions outperform simple sign-up bonuses, which give them a one time reward for participating.
\end{abstract}

\maketitle

\section{Introduction}

In the last decade, the sharing economy has upended a wide variety of industries, from bike sharing and car sharing to food delivery and hotels. However, despite the rapid rise and success of online sharing platforms, there is much about their design and operation yet to be understood.  Our lack of understanding is made clear by the fact that competing sharing platforms in the same space often have fundamentally different approaches to pricing, matching, subsidies, market positioning, etc.  The contrast between Uber, Lyft, and Didi with respect to pricing and market positioning in the context of ridesharing is a particularly salient example, e.g., \cite{uber_subsidy,uber_lyft_ola}.

As a result, the design and operation of sharing platforms has emerged as a lively and growing research area, with papers focused on efficiency in matching  \cite{stiglic2015benefits,alarabi2016demonstration}, pricing strategies \cite{benjaafar2015peer,banerjee2016multi}, equilibrium analysis \cite{fang2017prices,benjaafar2015peer}, empirical studies \cite{chen2015peeking,chen2016dynamic}, and more.

In this paper, \emph{our focus is on the task of designing subsidies, a.k.a., loyalty programs within sharing platforms.}  Subsidies play a crucial role in sharing platforms because they help to ensure that there is enough shared capacity in order to meet the demand.
For example, in the context of ridesharing, various forms of loyalty programs for drivers have been in place since the beginning of the industry, e.g., \cite{didi_subsidy,uber_subsidy}.
In addition to ensuring enough shared capacity is present in the system, loyalty programs play another important role in sharing platforms too.  They are   crucial tools when competing for and retaining sharing capacity.
For example, in the case of ridesharing, loyalty programs allow Uber and Lyft to lock-in drivers on one platform; thus both increasing the platform capacity and decreasing the capacity of competing platforms \cite{uber_incentives}.

The importance of these loyalty programs cannot be overstated -- maintaining capacity is the fundamental challenge of a sharing platform.  Thus, it is not surprising that they have become increasingly sophisticated as competition has heated up in the industry. In ridesharing, Uber and Lyft both began with simply offering a one-time reward, called a sign-up bonus, to new drivers. However,  loyalty programs have become considerably more complex and expensive as the industry has evolved.  Uber and Lyft both recently introduced loyalty programs to reward drivers after they reach certain trip milestones, i.e., drivers earn a ``power driver'' bonus after a minimum service requirement is reached. Indeed, platforms depend so much on these loyalty programs that they are willing to invest heavily in them. The cost of paying subsidies to boost supply and win market share is huge, e.g., Didi spends in excess of \$$4$ billion annually on subsidizing drivers \cite{didi_subsidy}.

Motivated by the importance of loyalty programs for sharing platforms, \emph{the goal of this paper is to study the optimal design of loyalty programs for sharing platforms and to understand the competitive advantage provided by moving toward more sophisticated loyalty programs.}

While we are motivated by ridesharing, our focus is on sharing platforms more broadly.  Thus, we build our work on classical models of two-sided platforms in economics \cite{rysman2009economics,weyl2010price,rochet2006two}.
Specifically, we first consider the case when the platform has monopoly in a  market with homogeneous product owners. We show that the optimal revenue can be achieved with a linear loyalty program (LLP) with a form similar to the ``power driver'' program used in practice.
We then consider the heterogeneous market.
In this case, however, finding the optimal loyalty program can be intractable as it requires searching over complex subsidy functions. Moreover, the resulting program may be impractical due to its complex structure. To tackle this problem, we prove that the optimal revenue can be achieved by a much simpler class of multi-threshold linear loyalty  programs (MTLP).  Moreover, compared to a general loyalty program which may be impractical, the form of MTLP facilitates implementation in practice.
This significantly reduces the required search space.
We further design a special MTLP program called a Hyperbolic Bonus (HB), which admits a simple hyperbolic bonus structure and is cost-efficient, in that it pays the necessary opportunity cost plus a constant regardless of the size of the market.

Our results highlight that the design of the optimal linear loyalty program has the property that it no longer extracts any commission fee when product owners have shared more than some required amount. This property mimics the design of Lyft's power driver program \cite{lyft_introduce_loyal}.

Further, we consider loyalty program design under competition.  Specifically, we study a setting in which two platforms compete for the scarce supply (shared capacity) with the objective of maximizing revenue.
Within this context, we conduct a comparative study of one-time sign-up bonuses and linear loyalty programs, in order to investigate the impact of the relative sophistication of loyalty programs on competition between the firms.
Our results show that sophisticated loyalty programs that reward suppliers via stepwise linear functions provide a meaningful competitive advantage over simple sign-up bonuses, which give them a one time reward for participating.


Moreover, we show that the heterogeneity of suppliers plays a critical role in deciding the optimal loyalty program used by platforms. This is important because suppliers in the sharing economy tend to exhibit significant heterogeneity in engagement.
For instance, it has been reported that there exist two different clusters of active drivers on Uber platform \cite{kooti2017analyzing},  where the highly active group members have nearly $6$ times longer driving time than low activity group members four months after their registration. This heterogeneity has a fundamental impact on platforms' strategies, both in a monopolistic or competitive market.
When a platform has monopoly, introducing a more complicated ladder-like progressive bonus can extract more revenue from  heterogeneous suppliers.

While in a competitive heterogeneous market, platforms compete through the design of loyalty programs and can  counter the ``winner take all'' phenomenon, which is known to be prominent in competition between platforms from classic two-sided market  theory, e.g., \cite{rysman2009economics}. Intuitively,   if users are homogeneous, platforms may try to attract all users and obtain higher revenue.  This results in a classic ``winner take all'' situation since, e.g., the more users a   platform has the more desirable it is for drivers, and vice versa.   On the other hand, if users are highly differentiated, the aggressive strategy of capturing the whole market can become too expensive. In this  case, it is wiser for platforms to focus only on distinct user groups. This creates an opportunity for firms to be strategic about market positioning and may yield situations where multiple platforms can survive, avoiding a ``winner take all'' scenario.

\textbf{Contributions of this paper.}
In summary, this paper makes the following contributions.

\emph{First}, this paper introduces a novel model to study subsidy program of a sharing platform, which is widely adopted in real world to overcome supply shortage.
Our model is built on classical two-sided market models, but it captures the fact that  suppliers can now benefit from using the resources for their own purposes.
It also captures supplier heterogeneity in self-usage values. Further, the model allows the study of competition between platforms.

\emph{Second}, we prove that a linear loyalty program (LLP) maximizes revenue over all possible loyalty programs in a market with homogeneous suppliers, and we  characterize the optimal linear loyalty program design.
For markets with heterogeneous suppliers, we establish that optimal revenue can be achieved using a set of multi-threshold linear loyalty programs (MTLP) that have a simple bonus structure and are implementation-friendly. We further exploit the structure of MTLP and design an MTLP called Hyperbolic Bonus (HB), which admits a very simple bonus form and is cost-efficient.

\emph{Third}, we further consider loyalty program design under competition. We conduct a comparative study of one-time  sign-up bonuses and linear loyalty programs, in order to investigate the impact of the relative sophistication of loyalty programs on competition between the firms. Our findings can be used to explain the fact that Uber and Lyft  initially introduced sign-up bonuses to attract driver registration, but have later shifted to using ladder-like linear loyalty programs \cite{lyft_introduce_loyal}.

\emph{Fourth}, we show that supplier heterogeneity has a dramatic impact on the market, e.g., on both the platform's pricing strategy and market positioning strategy.  Our findings can be used to explain why Uber and Lyft adopt different market positioning strategies to attract drivers and riders \cite{uber_lyft_ola,uber_lyft_compare}.

\emph{Finally}, to validate analytic results, we conduct a simulation using utility functions derived from Didi's transaction data, which is reported in \cite{fang2017prices}. Our results show that multi-threshold loyalty program achieves near-optimal revenue in heterogeneous markets, and that linear loyalty programs achieve higher revenue than sign-up bonuses, and they are more robust to supplier heterogeneity.

\textbf{Related literature.}
A large and growing literature studying the sharing economy has emerged in recent years.
These works focus on a wide variety of issues from efficiency in matching \cite{stiglic2015benefits,alarabi2016demonstration}, pricing strategies \cite{benjaafar2015peer,banerjee2016multi}, equilibrium analysis \cite{fang2017prices,benjaafar2015peer}, empirical studies \cite{chen2015peeking,chen2016dynamic}, and more.

At this point, existing empirical studies have been able to shed considerable light on the behavior of consumers in sharing platforms.  These studies inform the modeling choices we make in this paper.    For example, \cite{cohen2016using} estimates user surplus from demand data under different surge prices,  \cite{chen2015peeking} studies Uber's dynamic pricing strategies by collected API data from San Francisco and Manhattan, and shows that demand is much more sensitive to transaction prices compared to supply capacity.  Similar studies exist for other sharing platforms. In the case of Airbnb, \cite{lecuyer2017improving,quattrone2016benefits} study Airbnb's revenue and occupancy by crawling listing data from the website, and \cite{zervas2014rise} estimates the impact of Airbnb on the hotel industry.

The current paper fits into the growing literature that uses analytic models to study and optimize the design of sharing platforms, e.g.,
see \cite{rysman2009economics,weyl2010price,rochet2006two} and related references for an overview.  Such papers focus on two-sided market models where both sides in the market benefit from the participation of the other side, and platforms set prices to maximize  transaction volumes,  defined as the product of population sizes on both sides.  Within this literature, some papers focus specifically on the pricing and subsidy strategies for sharing platforms.  These are the most related to this paper.

Most such papers focus on optimizing the pricing strategy of a monopolistic platform.
For example, \cite{fang2017prices} shows that subsidizing product owners can substantially increase supply and, hence, increase social welfare. Similarly, \cite{benjaafar2015peer} studies the pricing strategies to achieve optimal revenue and social welfare.  Other highlights from the literature on pricing in a single platform are \cite{Banerjee:2015}, which shows that dynamic pricing is more robust than static pricing in the context of ride sharing; \cite{cachon2017role}, which shows that fixed ratio commission is nearly optimal for platforms and dynamic pricing helps increase social welfare; and \cite{banerjee2017pricing}, which provides an approach to approximately optimize revenue, throughput and welfare for a sharing platform based on a Markov model.

There is far less work studying competition between platforms.  The most prominent examples are \cite{rochet2003platform,armstrong2006competition},  which show that it is in the best interest of platforms to compete for users who are single-homing in order to attract loyal users.  These papers highlight the importance of considering supplier heterogeneity when studying competition between platforms, but leave open the question of how to design subsidies to attract and retain suppliers.

The importance of capturing user heterogeneity is also emphasized by \cite{banerjee2017segmenting}, which considers a model where buyers and sellers are segmented by their characteristics and proposes an algorithm for platforms to control the visibility among  segmented groups to achieve approximated optimal revenue.  The degree of heterogeneity is also an important feature in empirical studies, e.g., \cite{kooti2017analyzing} shows that Uber users are highly differentiated in various aspects.

In summary, there is a growing literature of analytic work studying pricing and subsidization in platforms.  This work has highlighted the importance of modeling consumer heterogeneity, and proposed explicit optimal subsidy strategy for practical use. It has also showed the performance of subsidy programs in the competition between two sharing platforms, but has only scratched the surface in understanding how subsidies and competition  impact  revenue of a sharing platform.  Our paper is the first to characterize optimal subsidization and to look at competition that is driven by subsidies.

\section{Model and Preliminaries} \label{sec:model}

The goal of this paper is to study the loyalty programs in sharing platforms, focusing on the most common case when supply is insufficient and loyalty programs are necessary to attract and retain supply.
To facilitate understanding,  we first describe our model for  a monopolistic market in this section. Then, we analyze this market in Section \ref{section:monopoly}. After that, we  extend the results to study competition between two platforms in Section \ref{section:competition}.
All the proofs of our results in this paper are in appendix.

\subsection{Demand and supply}
In our model, demand comes from product renters and supply comes from product owners.
The total demand is denoted by $D(q)$, where $q$ is the renting price that the platform charges for per unit demand.
For example, demand can be ride requests or apartment rental orders.
Assume that $D(q)$ is decreasing in $q$,  and  $q\in[q_{min},q_{max}]$.
%
In our model, demand is considered to be a function of price.
Quantitative case studies \cite{mohlmann2015collaborative, lamberton2012ours} have been conducted to show that economic benefit (maximize economic gain and save money) is the determinant of user's decision and behavior in the sharing economy, under different services such as bike sharing, car sharing and accommodation sharing.
Take the car hailing market for example, empirical studies show that renters are primarily sensitive to price \cite{chen2015peeking}, and wait time does not have major impact on passenger's behavior given real world ETA data \cite{fang2017prices}.
\begin{figure} 	
	\centering
	\includegraphics[scale=0.3]{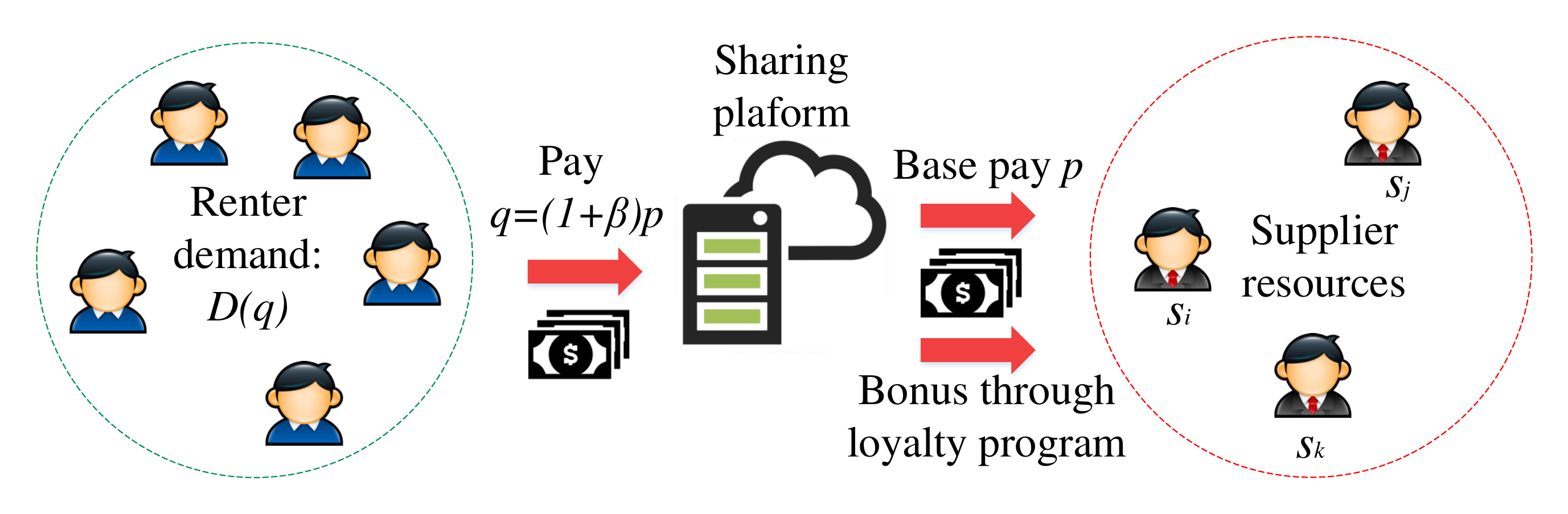}
	\caption{Platform charges renter $q$ for per unit demand, and pays suppliers $p$ for per unit sharing. Extra subsidies are paid to active suppliers to incentivize more sharing. }
	\vspace{-0.1in}
	\label{fig:sketch}
\end{figure}

In our paper, we will assume that $q$ is fixed, and focus on the optimal price and subsidy program on owners. The reason is that we consider the long-term average effect of pricing and user participation. This style of model has become popular in recent  years since it separates design issues associated with dynamics from the underlying economic competition between platforms, e.g., \cite{jiang2016collaborative,benjaafar2015peer}.
This assumption captures a platform's limitation in pricing in reality.
Though renter price fluctuates in real time (dynamic pricing), the long-term average is relatively stable. For instance, in the context of ride sharing, passengers are highly price sensitive \cite{fang2017prices}, so platform has to maintain a stable and reasonable ride price to attract and retain customers.
Further, the price of substitute services also limits the pricing power of the sharing platform, e.g., Uber tries to beat taxi fare a little bit in many cities \cite{uber_beat_taxi}. In fact, the passenger prices for UberX in London have stayed the same for the last two years, while prices in New York have only been changed once in the last two years.\footnote{See http://uberestimate.com/prices/New-York-City/all/ for price change log in New York City.}

We denote the set of suppliers  by $\mathcal{O}$. Since we focus on the case when supply is insufficient, we assume that  $D(q)>|\mathcal{O}|$ for $q\in[q_{min},q_{max}]$.
For each user $i\in \mathcal{O}$, we normalize his total resource to be  $1$ unit, and assume that he can divide the allocation of his resources into two parts: self-usage $x_i\in[0,1]$ and  sharing on platform $s_{i}\in[0,1]$.
Note that $x_i+s_{i}\leq1$.
Thus, the total supply on the platforms is:
\begin{equation*}
S=\sum_{i\in \mathcal{O}} s_{i}
\end{equation*}

To determine each owner's sharing level $s_i$,
 we model each product owner's utility as the sum of two different components: (i) utility derived from self-usage of the product, and (ii) income earned from sharing the product through the platform.

We denote $f_i(x_i)$ the self-usage utility. 
Note that each owner's $f_i(x_i)$ is not revealed to others, and it is defined on $[0,1]$.
In this paper, we assume that each owner's utility function $f_i(x_i)$ is concave increasing and differentiable on $[0,1]$.
This assumption guarantees full utilization of resources, i.e., $x_i+s_{i}=1$. Denote $p$ the per unit sharing price the platform offers to owners.
We also assume that $f'_i(1)=0$ and $f'_i(0)>p$. This ensures that self-usage is not a trivial option for product owners.
Therefore, the overall utility of owner $i$ is given by:
\begin{equation} \label{eq:utility_mono}
U_i=f_i(x_i)+p s_{i}.
\end{equation}
Each owner chooses his $x_i$ and $s_{i}$ values to maximize his  utility (\ref{eq:utility_mono}).
%

In the two-sided market literature, owner utility in (\ref{eq:utility_mono}) often takes the form  $U_i=f_i(x_i)+p\min\{1, D/S\}s_{i}$, to take into account product utilization, e.g.,  \cite{cachon2017role, Even-dar:2009,fang2017prices}.
In our case, since optimal loyalty programs are primarily needed when supply is insufficient, we assume without loss of generality that $\min\{1, D/S\}=1$ in this paper.\footnote{For the situation when supply is sufficient, there is no need for loyalty programs or other promotions for suppliers. Platforms can achieve maximum revenue by choosing optimal prices \cite{fang2017prices,benjaafar2015peer}.}

%

\subsection{Platform price and revenue}


In practice, platforms usually extract a  commission fee, i.e., a fixed  percentage of each ride's value. We define the commission fee $\beta\geq 0$ and conveniently write
\begin{equation}
q = (1+\beta)\cdot p
\end{equation}
Recall that $q$ is the service price to renters and $p$ is the payment to owners for per unit resource sharing.
Revenue is the profit from the gap of the income from renters and payment to owners.

However, supply shortage has been shown to prevent a platform from obtaining the maximum revenue since many renter orders are not fulfilled \cite{fang2017prices}.
Thus, in practice, a common strategy used by sharing platforms  to overcome this challenge is to subsidize resource providers. Yet, despite the wide adoption of subsidy programs, it still remains to understand how to design an optimal subsidy program.

To carry out our study on optimal subsidies, we first introduce a general definition for a subsidy program. A \emph{subsidy program} gives owner $i$ a bonus $W(s_i)\geq0$ according to his sharing level $s_i$. Note that $W(s_i)$ is a non-decreasing function. It is clear that a subsidy program impacts a platform's revenue, since the platform has to pay its suppliers an additional amount for their sharing.
To take this into account when analyzing platform revenue, we define platform's revenue as:
\begin{equation}
R= (q-p) S - \sum_{i}W(s_i),\label{eq:revenue_loyal}
\end{equation}
where $\sum_{i}W(s_i)$ is the total subsidies paid to owners. For notational convenience, we define $B(s)$ to be the marginal bonus, i.e.,
\begin{equation}
	W(s_i)=\int_{0}^{s_i}B(s)\ud s. \label{eq:B-def}
\end{equation}

\subsection{Heterogeneous suppliers}
Heterogeneity is a fundamental issue for sharing platforms, as we have highlighted in the introduction. To model heterogeneous suppliers, we consider $n$ different classes of owners in the market with self-usage utility functions  $f_i(x_i)$, $i\in\{1,2,...,n\}$. To simplify the analysis, we represent each class as one owner, and use owner $i$ to refer to owners in class $i$ for simplicity.  This is for convenience and is without loss of generality. All of our results can be extended to the case when there are arbitrary number of owners in each class.
We assume that any two classes of users are highly differentiated. Specifically, we assume that  $f_1'(x)> f_2'(x)>...>f_n'(x)$, for $x\in[0,1]$.  This assumption is natural given that empirical studies have shown marked differentiation between users \cite{kooti2017analyzing}.


%

\section{Subsidies in a Monopolistic Market}\label{section:monopoly}
We begin our analysis with the study of a monopolistic platform.
Our main result in this section is that maximal revenue can be achieved in both homogeneous and heterogeneous markets.  In the homogeneous market we show that linear loyalty programs are optimal while in heterogeneous markets we show that multi-level threshold programs are optimal.  Finally, we also introduce a novel form of subsidies, hyperbolic bonuses, that are cost-efficient and easy to determine.

\subsection{Homogeneous markets} \label{subsec:homo}

In this subsection we show that linear loyalty programs achieve maximal revenue across any general forms of loyalty program when suppliers are homogeneous. Linear loyalty programs are defined as follows.

\begin{definition} \emph{ A \textbf{linear loyalty program (LLP)} is a subsidy program under which  the platform pays owners an extra bonus $B$ for per unit sharing, in addition to the base pay $p$, to those who share higher than some threshold $t$. Under a linear loyalty program, the utility of owner $i$ is:
\begin{equation} \label{eq:utility-lp}
U_i=f_i(x_i)+p s_{i}+ B(s_{i} - t)^+,
\end{equation}
where $(s_{i} - t)^+=\max\{s_{i} - t,0\}$.}
\end{definition}

The following theorem summarizes the optimal bonus volume, service threshold and revenue obtained under linear loyalty programs. 

\begin{theorem}\label{thm:lp_mono}
Given homogeneous owners, i.e., $f_i(x)=f(x)$ for all $i\in\mathcal{O}$, and platform's prices on supply and demand, i.e., $p$ and $q$, the following loyalty program achieves the maximum revenue over all subsidy programs in a monopoly market:
	\begin{align}
		B^*&=q-p,\nonumber\\
		t^*&=s-\frac{f(1-s_0)-f(1-s)-p(s-s_0)}{q - p}.
	\end{align}
	Here $s$ and $s_0$ satisfying $f'(1-s)=q$ and $f'(1-s_0)=p$ are the owner  sharing levels with and without the loyalty program, respectively.
	Moreover, the maximum revenue that the platform obtains is:
	\begin{align}
	R^*=B^*\cdot t^*.
	\end{align}
\end{theorem}

The meaning of above $B^*$ and $t^*$ is that the platform should boost supply by \emph{returning all revenue to owners who meet the minimum sharing requirement},  e.g., extract no commission fee for drivers who have reached a certain trip milestone on the platform.  This parallels the current design of  Lyft's power driver program \cite{lyft_introduce_loyal}.

The theorem is proved in two steps. The first step is to show the optimality of the linear loyalty program among all possible forms of loyalty program. We prove that linear loyalty program at least achieves the same revenue as any general loyalty program. The second step is to compute the optimal bonus, which turns out to be exactly the price gap between $q$ and $p$. Interestingly, the optimal bonus $B^*=q-p$ is independent of the form of function $f(x)$.

%
%

The intuition behind this surprisingly simple form of the optimal subsidy program is that, this linear form of loyalty program can exactly cover owner's opportunity cost for one to increase sharing amount, i.e., barely compensate for the lost of self-usage benefit, in order to maximize platform revenue.

\begin{figure} 	
	\centering
	\includegraphics[scale=0.28]{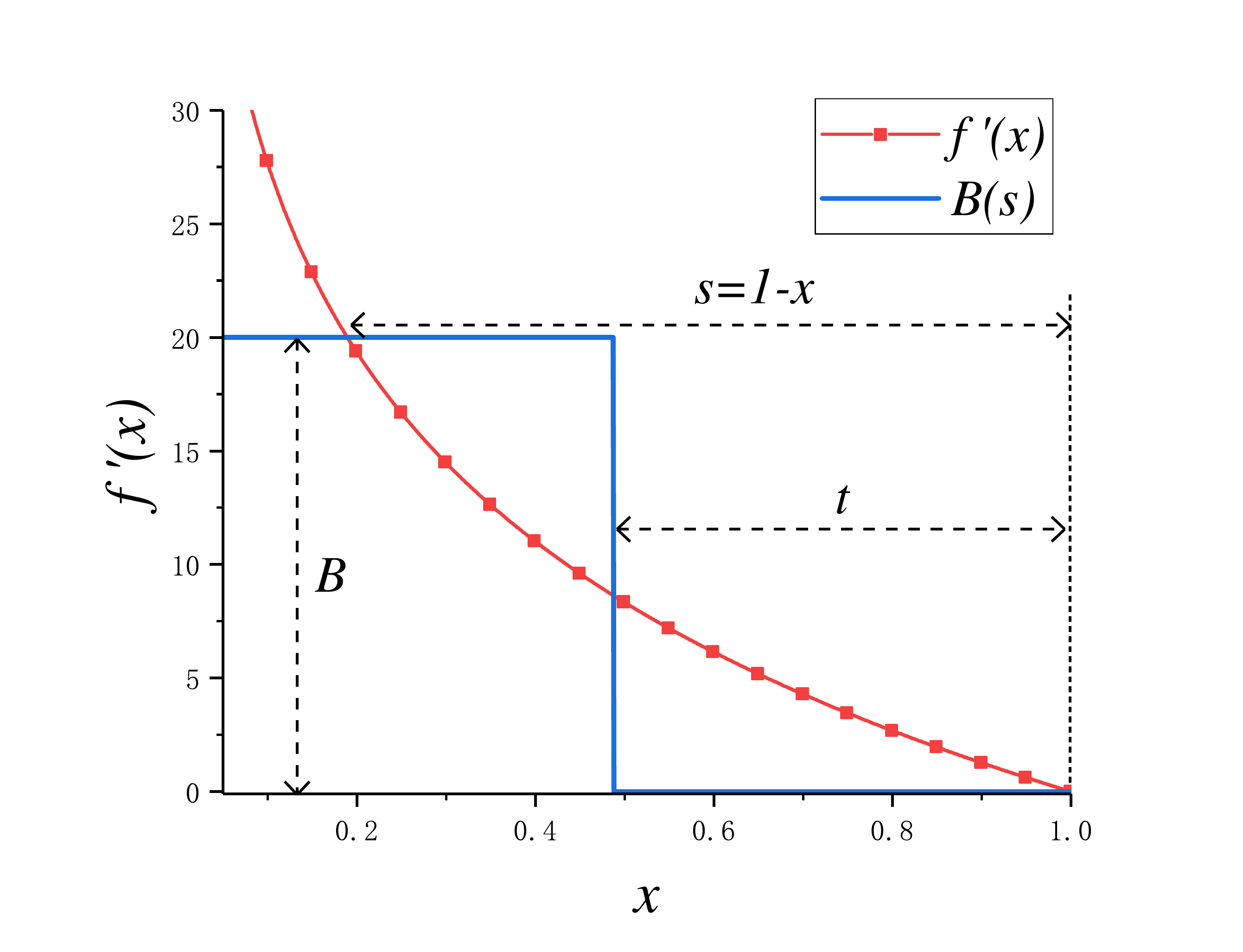}
	\vspace{-0.2in}
	\caption{Optimal linear loyalty program pays owners $B=q$ for per unit sharing after his sharing level exceeds threshold $t$. Owner's optimal sharing $s$ will be at the intersection point of $f'(x)=B$, where the marginal benefits of self-usage and sharing are equal. }
	\vspace{-0.1in}
	\label{fig:single_instruction}
\end{figure}

With Theorem \ref{thm:lp_mono}, we can further compute the optimal supply price $p$ and the corresponding platform revenue below.
\begin{corollary}\label{thm:opt_mono}
	Given homogeneous owners and fixed demand price $q$, the maximum platform revenue is achieved by letting
	\begin{align}
		p&=0, \nonumber\\
		B&=q, \nonumber\\
		t&=s-\frac{1}{q}(f(1)-f(1-s)),
	\end{align}
	where owner's sharing level $s$ satisfies $f'(1-s)=q$.
\end{corollary}
Interestingly, this Corollary \ref{thm:opt_mono} shows that the optimal supply price $p$ is zero.  That is to say, it is the platform's optimal strategy to pay owners nothing unless they share more than some given threshold. From the above, we see that the platform should pay owners exactly the amount to  cover their  opportunity cost of using the resources themselves (users will start to share only when the marginal benefit of self-usage is lower than $p+B$). Therefore, setting  $p=0$ gives the minimum opportunity cost of owners.
Figure \ref{fig:single_instruction} shows an optimal linear loyalty program. 

Note that this result extends to the case when owners have linear usage/sharing cost, e.g., gas, taxation \cite{fang2016www}.  In this case, the optimal owner price $p$ in Corollary \ref{thm:opt_mono} should be the price that exactly equals the cost per unit sharing, leaving no extra revenue for owners who share less than the threshold $t$.

\subsection{Heterogeneous markets}

We now turn to the case when owners are heterogeneous. Recall that for the $n$ owners, we have $f_1'(x)> f_2'(x)>...>f_n'(x)$, for $x\in[0,1]$.
We first introduce a result describing the owners' participation in loyalty program, given the marginal bonus $B(s)$ non-decreasing.

\begin{proposition}\label{prop:non-decreasing}
Under a loyalty program with non-decreasing marginal bonus $B(s)$, if owner $i$ benefits from participating in the loyalty program, i.e., he increases his sharing level compared to the amount without loyalty program, all owners $j>i$ will also participate in the loyalty program $B(s)$ and boost sharing.
\end{proposition}
The intuition of the above result is that if owner $i$ thinks that it is more profitable to share more, other owners $j>i$ will also increase sharing since they have even lower self-usage benefits and hence lower opportunity costs.
Thus, the platform can choose a loyalty program such that all owners $j>i_0$ participate. Without loss of generality, we only consider loyalty programs that all $n$ owners will participate in this paper to simplify the presentation.\footnote{Our results can easily be extended to the case when only owners in $\{i_0, ..., n\}$ participate in the program.}

We are interested in finding the optimal loyalty program for this heterogeneous case.
It is natural to extend the single threshold in the linear loyalty program to a $n$-threshold version with a progressive bonus, as follows.

\begin{definition}
\emph{A \textbf{multi-threshold loyalty program (MTLP)} is an $n$-threshold linear loyalty program, where the corresponding progressive bonus for the sharing amount $s\in[t_{k-1},t_{k})$, is $B_k$, for $B_1<B_2<...<B_n$ and thresholds  $0=t_0<t_{1}<t_{2}<...<t_{n}<t_{n+1}=1$.}
\end{definition}

Under MTLP, owner $i$ with sharing level $s_i\in[t_k,t_k+1)$ has utility:
\begin{align}\label{eq:multi_utility}
	U_i=f_i(x_i)+p s_{i}+ \sum_{j=1}^{k-1} B_j(t_{j+1} - t_{j})+ B_k(s_{i}-t_{k}).
\end{align}
Here $\sum_{j=1}^{k-1} B_j(t_{j+1}- t_{j})$ is the aggregate bonus for sharing  up to $t_k$, and $B_k(s_{i}-t_{k})$ is additional bonus at rate $B_k$. 

As in the homogeneous market case, it can be shown that the optimal owner price is $p=0$, since any $p>0$ is essentially given to users for free, when their sharing levels are below the point where $f'_i(x)=p$.
Thus, by using $p=0$, the platform's revenue is:
\begin{equation}\label{eq:rev_multi}
R=q\sum_{i=1}^{n}s_i -\sum_{i=1}^{n}\big[B_{k_i}(s_{i}-t_{k_i})+\sum_{j=1}^{k_i-1}B_{j}(t_{j+1}-t_{j})\big].
\end{equation}
where $s_i\in[t_{k_i},t_{k_{i+1}})$.
The first term of (\ref{eq:rev_multi}) is the income from renters, and the second term is the bonus paid to owners.

\begin{theorem}\label{thm:multi_opt}
MTLP achieves optimal revenue among all loyalty programs with non-decreasing marginal bonus $B(s)$.
\end{theorem}

Theorem 3.6 is proven by showing that for any loyalty program with a non-decreasing B(s), there exists an MTLP counterpart that can achieve the same revenue. Therefore, it is sufficient for a platform to optimize over ladder-like bonus programs with multiple thresholds for achieving optimal revenue. This significantly reduces the search space, and the resulting subsidy programs are also highly practical.

Note that given $p=0$ and the owner utility function in (\ref{eq:multi_utility}), owner $i$'s optimal strategy $(x_i,s_i)$ should guarantee that the marginal incomes of sharing and self-usage are equal. In other words, a reasonable sharing choice for owner $i$   should belong to  the intersections of the two curves, i.e., where $f'_i(x)=B_j$.
We denote $s_{i,i-1}$ the sharing point of owner $i$ at the intersection of $f'_i(x)$ and $B_{i-1}$ (not necessarily his final choice), i.e., we have $f'_i(1-s_{i,i-1})=B_{i-1}$.


Below, we construct a particular MTLP and show that it is cost-efficient. To do so, we first derive a necessary condition that any multi-threshold loyalty program must satisfy.

\begin{theorem}\label{thm:multi_necessary}
	For any MTLP, we have for all $i=1,2,...,n$ that:
	\begin{align}\label{eq:multi_constraint}
           B_i(s_i-t_i)+B_{i-1}(t_{i}-s_{i,i-1})\geq \int_{1-s_{i}}^{1-s_{i,i-1}}f_i'(x)\ud x.
\end{align}
\end{theorem}
Condition in (\ref{eq:multi_constraint}) essentially means that any  MTLP guarantees that the bonus owner $i$ receives over the interval $[s_{i, i-1}, s_i]$ should at least cover his  self-usage opportunity cost in the same interval.

Now we are ready to present  a simple MTLP called Hyperbolic Bonus \textbf{(HB)}. The advantage of HB is that it admits an exact characterization of the bonus values. The intuition beyond HB is to tighten the subsidies to owners, by making the left hand side and the right hand side of (\ref{eq:multi_constraint}) equal for all owners. HB actually exploits the power of differential pricing, so that owner $i$'s best strategy satisfies  $s_i\in[t_{i},t_{i+1})$, i.e., each owner's sharing level resides in one corresponding interval. The exact form of HB is as follows.

\begin{definition} \emph{The \textbf{Hyperbolic bonus (HB)} program chooses $B_i$ as
\begin{equation}\label{eq:multi_bonus}
	B_i=\frac{q}{n-i+1}, i=1,2,...,n.
\end{equation}
The corresponding thresholds $\{t_i, i=1, ...,n\}$ satisfy (\ref{eq:multi_constraint}) with equality, i.e., for all $i$,
\begin{align}\label{eq:multi_constraint_eq}
           B_i(s_i-t_i)+B_{i-1}(t_{i}-s_{i,i-1}) =  \int_{1-s_{i}}^{1-s_{i,i-1}}f_i'(x)\ud x.
\end{align}
}
\end{definition}

\begin{figure}[!t] 	
	\centering
	\includegraphics[scale=0.28]{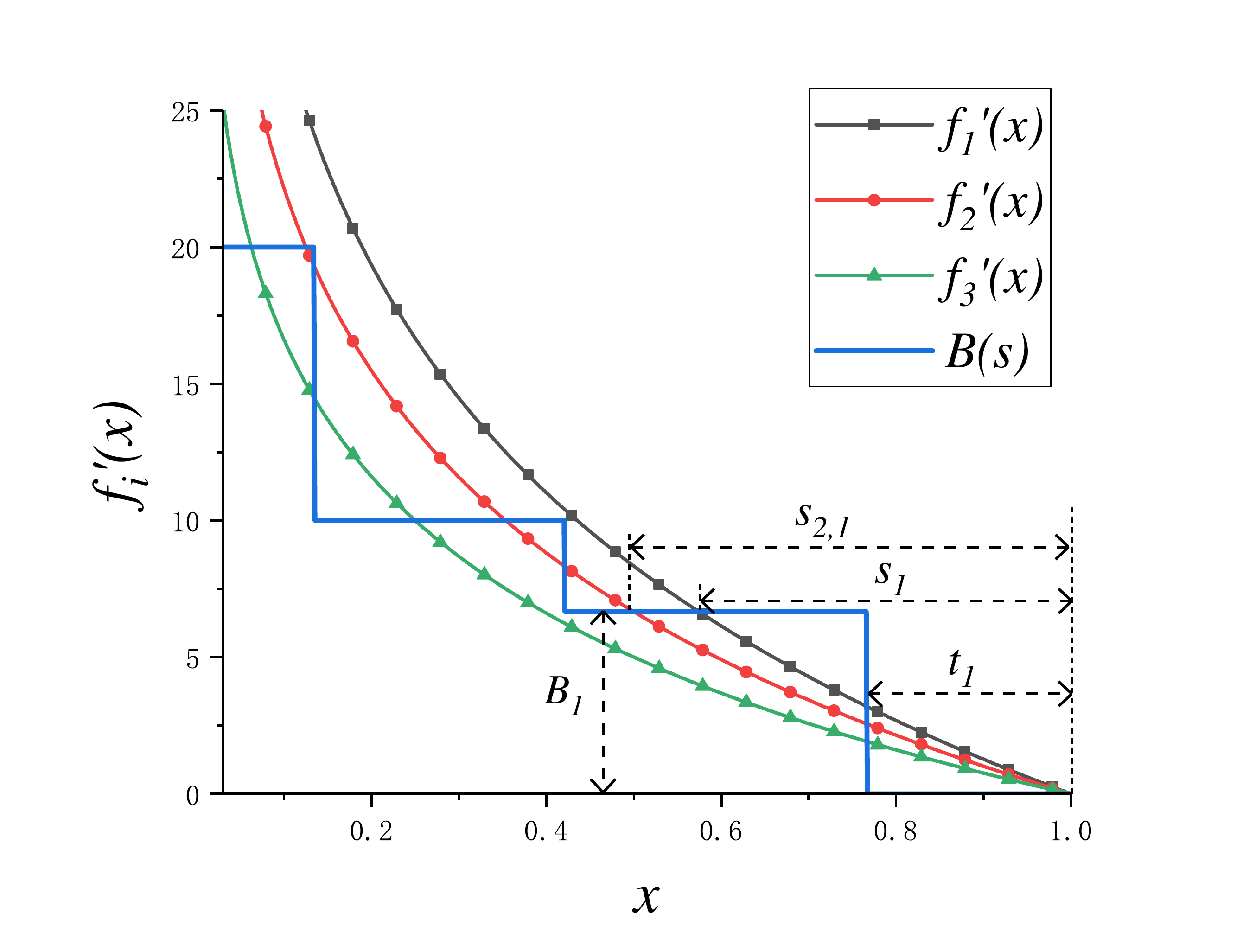}
	\vspace{-0.2in}
	\caption{A multi-threshold loyalty program introduces ladder-like bonus which rewards owners progressively. Here owner $1$ shares at the intersection point of $f'_1(1-s_1)=f'_1(x)=B_1$ and $s_{2,1}$ is the intersection point where $f'_2(x)=B_1$.  }
	\vspace{-0.1in}
	\label{fig:multi_instruction}
\end{figure}

The existence of such $t_i$ in HB is shown in the proof of Theorem \ref{thm:multi_opt}. Figure \ref{fig:multi_instruction} shows an example of a multi-threshold loyalty  program and owners' derivatives in the case of $n=3$.

Essentially the hyperbolic bonus at each stage is inversely proportional to the number of owners who earn bonus in this stage. Under such bonus strategy, platform pays little amount of reward for sharing levels that many owners are capable to reach.
The following theorem shows explicitly that HB is cost-efficient.

\begin{theorem}\label{thm:hb}
The amount of over-subsidization under the HB program is upper bounded by $q$, i.e.,
\begin{eqnarray}
\sum_{i=1}^{n} \bigg[W_i(s_{i}^{HB}) - \int_{1-s_{i}^{HB}}^1f'_i(x_i)\ud x \bigg]\leq q. 	\label{eq:hb-gap}
\end{eqnarray}
Here $s_{i}^{HB}$ is owner $i$'s sharing level under HB, $\sum_{i=1}^{n} W_i(s_{i}^{HB})$ is the overall subsidization given by HB and $\int_{1-s_{i}^{HB}}^1f'_i(x_i)\ud x$ denotes the opportunity cost of owner $i$.
\end{theorem}
Since any subsidy program has to at least cover the opportunity costs of owners, (\ref{eq:hb-gap}) shows that HB almost does not provide any additional subsidization, i.e.,   the extra payment  is at most $q$, which is \emph{independent of the market size $n$}.
As a platform's profit typically increases with $n$, a fixed revenue gap $q$ also implies that our algorithm is \emph{cost-efficient}.
Moreover, as we will show in Section \ref{subsection:homo-market}, in addition to its efficiency, HB also achieves very good revenue performance.


\section{Subsidies in a Competitive Market}\label{section:competition}
In this section, we consider a market with two platforms, and conduct a comparative study on competition through loyalty programs.
In this case, let $q_a, p_a$ and $q_b,p_b$ denote the renter price and owner reward on platform $a$ and $b$, respectively.
The following results discuss only the single-threshold linear loyalty program, showing that when facing a typical heterogeneous market with owners from $2$ groups, even a single-threshold program helps company compete.

Specifically, consider two groups of owners ($n=2$)  in the market, with  $f'_1(x)>f'_2(x)$ for $x\in[0,1]$. Note that $n=2$ is a practical situation, e.g., users can often be categorized as active or inactive \cite{kooti2017analyzing}. We assume that demand on a platform is a function of renter prices from both platforms, and supply is insufficient, i.e., $D_j(q_a,q_b)>|\mathcal{O}|, j=a,b$ for $q_a, q_b\in[q_{min},q_{max}]$.
Then,  the utility of owner $i$ is:
\begin{equation} \label{eq:utility}
	U_i=f_i(x_i)+p_a s_{ia}+p_b s_{ib},
\end{equation}
where $s_{ia}$ is owner $i$'s sharing level on platform $a$, and $s_{ib}$ is his sharing on platform $b$.
In the case when $p_a=p_b$, we assume that owners will share equally on both platforms.

To demonstrate the performance of linear loyalty program, we also introduce another widely adopted loyalty program,
\emph{sign-up bonus}, for comparison. To implement a sign-up bonus, the platform pays a one-time reward $B$ for owners who share \emph{exclusively} on the platform,  i.e., if an owner decides to share on this platform and receives a sign-up bonus, he cannot share on other platform simultaneously.
For example, if platform $a$ introduces sign-up bonus and owner $i$ joins this program, his utility will be
\begin{equation} \label{eq:utility-sb}
U_{i}=f_i(x_i)+p_a s_{ia}+B_a,
\end{equation}
where $B_a$ is the sign-up bonus from platform $a$.
In short, sign-up bonus is a one time payment to lock-in owners in competition.

Therefore, each platform can have three possible situations in the following discussion: adopting no loyalty program, adopting a linear loyalty program and adopting a sign-up bonus.

In this section, we assume that both platforms adopt the same commission rate, i.e., $\beta_a=\beta_b=\beta$.
The reasons why we consider identical $\beta$ are two fold.
The first is that competition in real world shows that platforms usually adopt the same commission rate, e.g., Uber and Lyft may vary their prices in different time and places, but they both extract a fixed $25\%$ commission rate.
The second is that an identical commission rate can help define the "stronger" and "weaker" platforms in our comparison.
With identical $\beta$, the platform that charges a higher renter price $q$ will also have a higher owner price $p$ ( recall that $q=(1+\beta)p$ ). We say that  this platform is "stronger" and more attractive to users than the other platform in competition.
For example, this platform may be better in service, reputation, and efficiency in matching demand and supply, etc., so that customers  want to pay for its higher price.
Note that the asymmetry in customer attractions and prices are common in real world: a less competitive manufacturer has to charge a lower price to win customers \cite{fang2016market}, sacrificing its revenue.
Naturally, we will show how a loyalty program can help the "weaker" platform to win market share.

\subsection{Competition without loyalty program}
We first provide results in a competition without loyalty programs as a baseline.
Here, there are two distinct cases to consider: (i) when the market is symmetric ($p_a=p_b$) and (ii) when the market is asymmetric ($p_a<p_b$).
\begin{proposition}\label{prop:no_loyal}
When both platforms adopt the same commission rate and neither of two platforms adopt loyalty programs, if $p_a=p_b$, platforms will equally share the supply, i.e., $S_a=S_b$. Otherwise, the platform with a lower supplier price has zero revenue.
\end{proposition}
In the symmetric case, the market is well-behaved, i.e., both platforms can win some market share.
In contrast, in the asymmetric case, things are different.  In particular, the asymmetric case highlights that market is quite fragile and, as soon as any asymmetry exists then the market becomes ``winner-take-all.''   This is a common phenomenon in two-sided markets and is a result of the fact that more supply makes the market more desirable for demand, and vice versa.

For our purposes, this result highlights the need for loyalty programs.  A ``weaker'' firm (one with smaller $p$), must find a way to compete in order to stay viable, and loyalty programs provide such an opportunity. Note that we only consider the situation when supplier price of the ``stronger'' firm is less than its competitor's demand price, i.e., if platform $b$ is ``stronger'' ($p_b>p_a$), we must have $p_b<q_a$. Otherwise, platform $a$ will never be able to attract owners, since platform $b$ is offering an owner reward even higher than platform $a$'s income from renters.
This condition is natural and common in practice, since platforms do not tend to coexist if one is extremely more efficient.

\subsection{Loyalty program on one platform}
The results in the case of no loyalty programs highlight the need for a ``weaker'' firm to compete via the introduction of a loyalty program. The results in this section highlight that such an approach to competition can be successful.  However, our results also highlight that there is an ``arms race''  underlying competition via loyalty programs.  As the ``weaker'' firm adopts a loyalty program, the ``stronger'' firm also has incentives to adopt a loyalty program.  However, our results also highlight that consumer heterogeneity eases the ``winner take all'' nature of this competition.


We denote $s_{1a}$, $s_{1b}$ as the sharing level of owner $1$ when he choose to share on platform $a$ or $b$, respectively. 
Denote $\lambda= \frac{s_{1a}}{s_{1b}}$. We always have $\lambda<1$, since $p_b>p_a$ and owner $1$ will share more on platform $b$.

To begin, we first consider the case where the ``weaker'' firm adopts a simple loyalty program -- a sign-up bonus -- in order to compete.  
The following theorem highlights that this approach can be successful.

\begin{theorem} \label{thm:sb}
Suppose $p_a<p_b<p_a(1+ \lambda \beta)$ and $\beta_a=\beta_b$.  Further, suppose platform $b$ does not use a loyalty program.
Then, there exists a sign-up bonus for platform $a$ under which platform $a$ has a strictly positive revenue.
\end{theorem}


Our proof in appendix provides a more detailed characterization of the exact form of the sign-up bonus and the revenue for each platforms.
We also observe in the proof that the resulting revenue of the firms depend heavily on whether firm $a$ tries to attract both users, or just one user. If the owners are diverse enough, the two firms will settle at an equilibrium where each attracts one owner (hence, both survive in the competition).
Otherwise when owners are more alike,  firm $a$ will try to take away all users (See Fig. \ref{fig:lp_k} and Fig. \ref{fig:critical_k} in Section \ref{subsectiobn:compete}  for experimental results).

\begin{figure*}[ht]
	\begin{minipage}[t]{0.33\linewidth}
		\centering
		\includegraphics[scale=0.23]{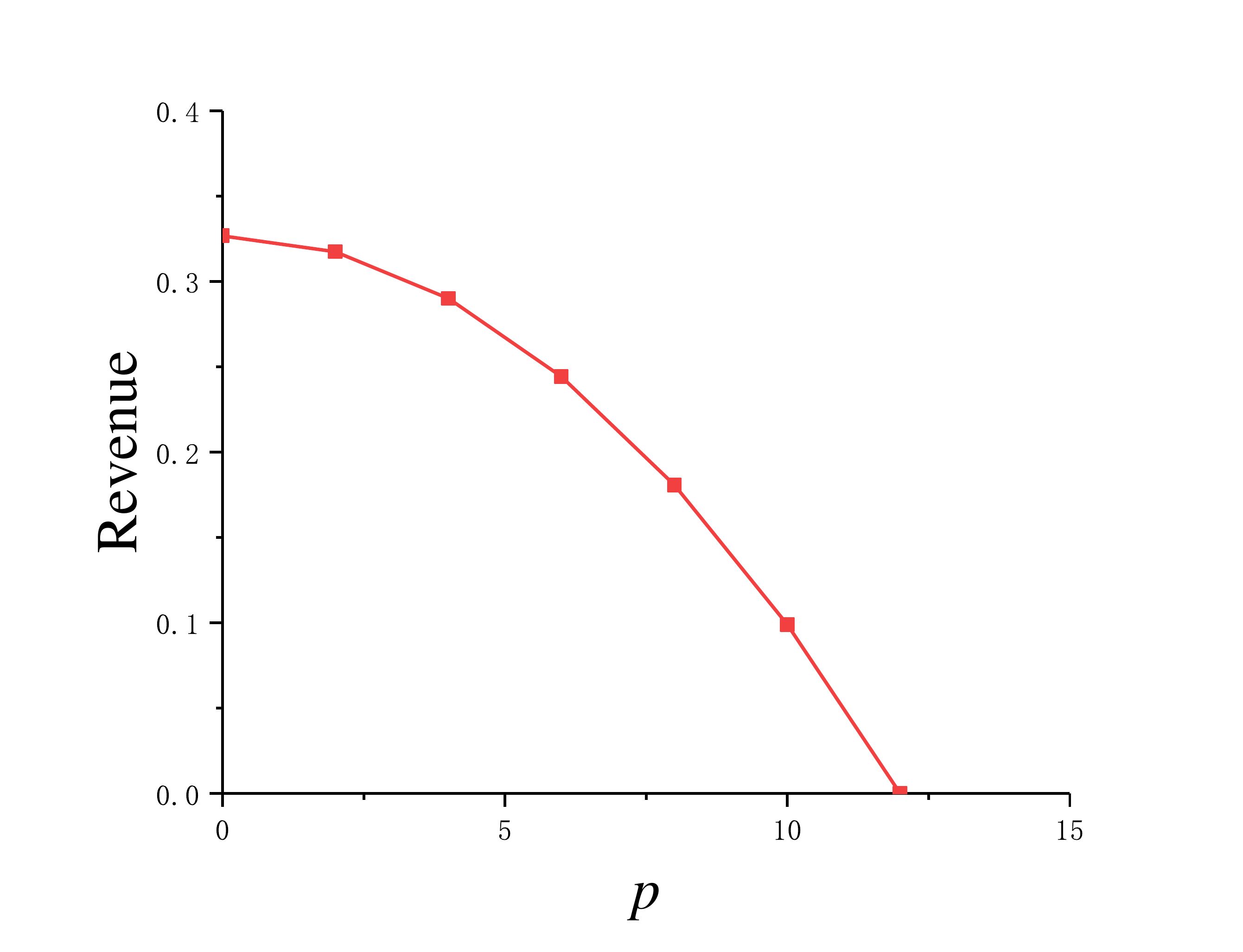}
		\vspace{-0.2in}
		\captionsetup{width=.9\linewidth}
		\caption{Platform's revenue decreases in supplier price $p$ under fixed renter charge $q=12$ and optimal linear loyalty program in Theorem \ref{thm:lp_mono}. It is optimal for platform to pay supplier zero base pay, i.e., $p=0$.}
		\label{fig:monopoly}
	\end{minipage}%
	\begin{minipage}[t]{0.33\linewidth}
		\centering
		\includegraphics[scale=0.23]{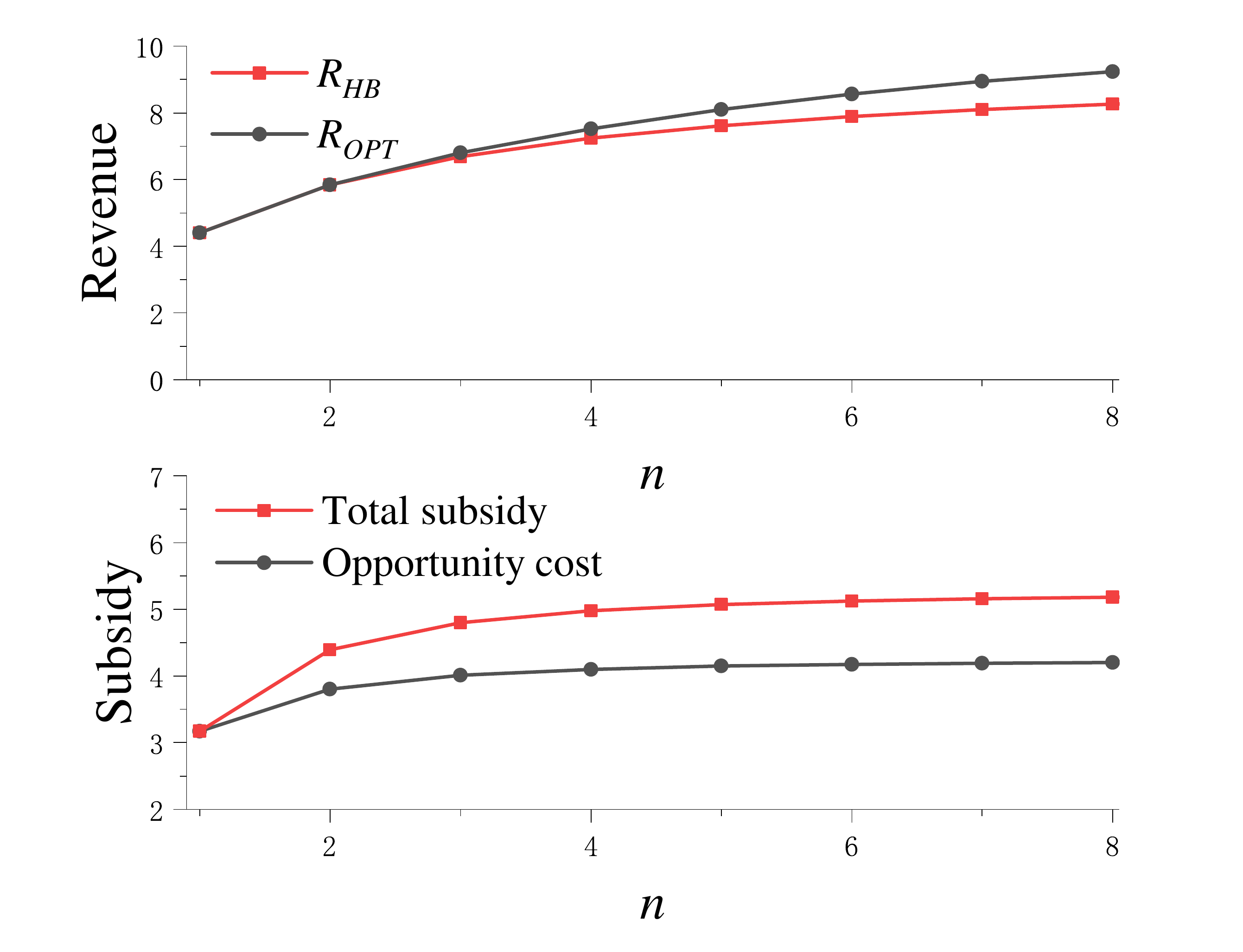}
		\vspace{-0.2in}
		\captionsetup{width=.9\linewidth}
		\caption{Revenue under the Hyperbolic bonus (HB) program $R_{HB}$ is close to the optimal revenue $R_{OPT}$, and the total subsidy is a constant beyond opportunity cost, independent of $n$ ($q=12$). }
		\label{fig:multi_revenue}
	\end{minipage}
	\begin{minipage}[t]{0.33\linewidth}
		\centering
		\includegraphics[scale=0.23]{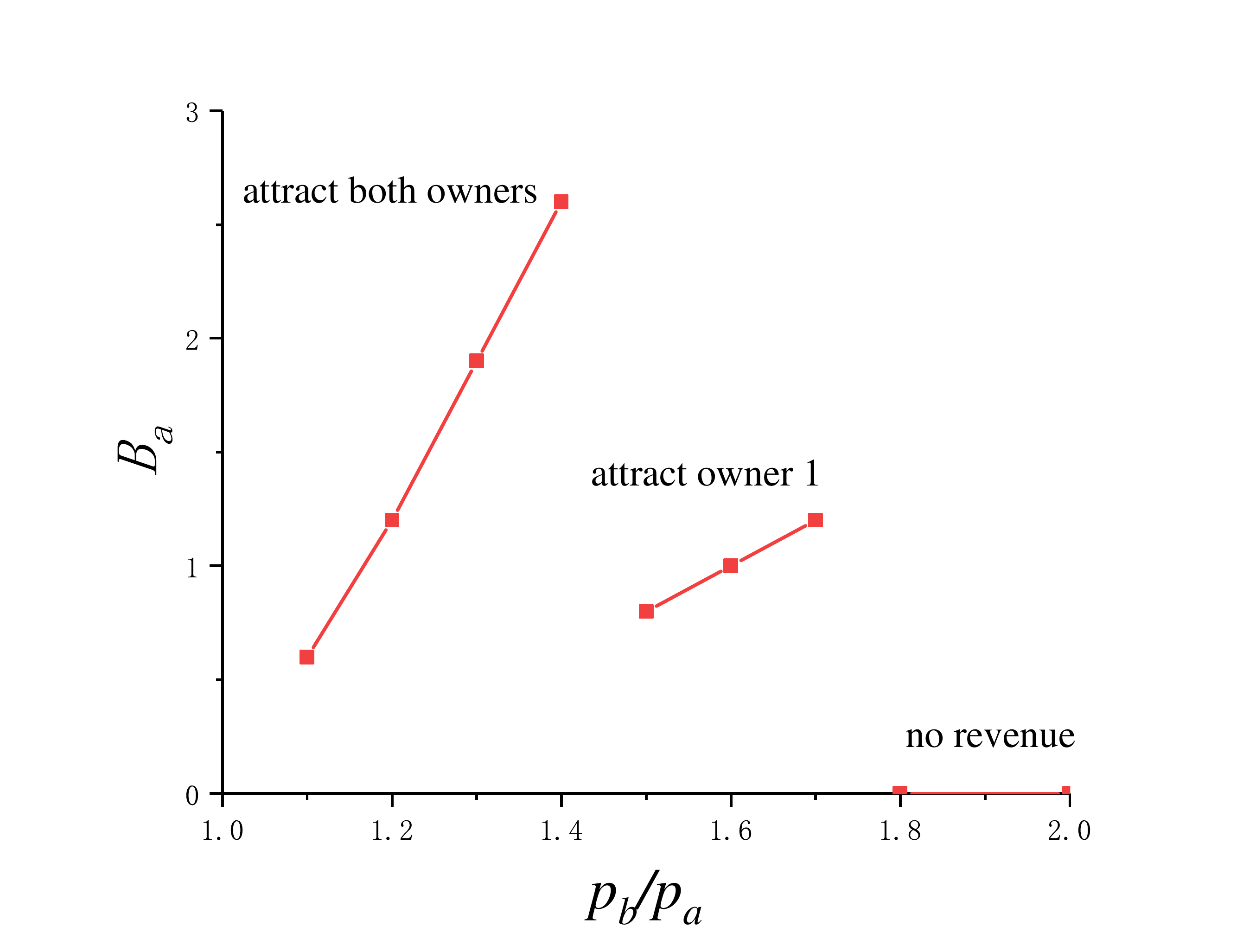}
		\vspace{-0.2in}
		\captionsetup{width=.9\linewidth}
		\caption{Platform $a$'s optimal sign-up bonus in three regimes: (i) attracting both owners when $p_b/p_a$ is small, (ii) attracting only owner $1$ when $p_b/p_a$ becomes larger, and (iii) no revenue when $p_b/p_a$ is large ($p_a=10$).}
		\label{fig:sb_p2}
	\end{minipage}
	\vspace{-0.2in}
\end{figure*}

Importantly, notice that it is not always possible for platform $a$ to use a sign-up bonus to compete.
In fact, we see from Theorem \ref{thm:sb} that if $p_b$ is very close to $q_a$, a sign-up bonus cannot ensure a positive revenue for platform $a$ (see Figure \ref{fig:sb_p2} for experimental result).
On the other hand, the theorem also highlights that platform $a$ can obtain a  positive revenue if $p_b$ is not too much larger than $p_a$.

Our next result highlights that introducing the more sophisticated linear loyalty program can also allow the ``weaker'' firm to gain   market share.

\begin{theorem}\label{thm:lp}
Suppose $p_a<p_b<q_a$, $\beta_1=\beta_2$, and platform $b$ does not use a loyalty program. Then, there exists a linear loyalty program $(B_a,t_a)$ for platform $a$ under which  platform $a$ can achieve a positive revenue.
\end{theorem}


Interestingly, our proof highlights that the optimal linear loyalty program to attract a specific class of owner turns out to be the same as in the monopolistic market (see Theorem \ref{thm:lp_mono}).  It gives a bonus that is equal to the platform's commission fee, i.e.,
\begin{equation}
	B_a=q_a-p_a.
\end{equation}

Importantly, the increased sophistication of the linear loyalty program as compared to the sign-up bonus means that platform $a$ can compete even if it is much weaker, as long as $p_b<q_a$ (sign-up bonus requires $p_b<p_a(1+\lambda \beta)$).
This highlights the value of increased sophistication.

Similar to Theorem \ref{thm:sb}, under a linear program, firms will also be strategic about which owner group to attract, and the results also depend largely on the heterogeneity of owners (See Fig. \ref{fig:lp_k} and Fig. \ref{fig:critical_k} in Section \ref{subsectiobn:compete} for experimental results).

\subsection{Loyalty programs on both platforms}
In fact the increased value of increased sophistication shows up even more prominently when sign-up bonus and linear loyalty programs are compared directly.


To see this, consider competition between two identical platforms with different forms of loyalty programs: platform $a$ uses a linear loyalty program and platform $b$ uses a sign-up bonus.  In this case, the platform using a linear loyalty program can always squeeze out its competitor while maintaining a strictly positive revenue.

\begin{theorem}\label{thm:vs}
Suppose $p_a=p_b=p$ and $q_a=q_b=q$.  If platform $a$ adopts a linear loyalty program and platform $b$ adopts a sign-up bonus, then platform $a$ can always maintain a strictly positive revenue while making platform $b$ gain zero revenue, i.e., the platform with linear loyalty program can force the competitor with sign-up bonus out of market.
\end{theorem}

This result highlights that if one platform adopts a sophisticated linear loyalty program, it is crucial for competing platforms to follow suit quickly. In fact, exactly that was observed in the case of ridesharing platforms, where both Uber and Lyft moved from sign-up bonuses to linear loyalty programs at nearly exactly the same time \cite{uber_copycat_loyalty}.
The intuition behind the fact that linear programs are more powerful is that they boost supply by incentivizing sharing, while sign-up bonuses only encourage participation, even under the strong assumption of exclusiveness.

%
%

\section{Case Study}\label{subsec:case_didi}

We conclude the paper by illustrating the results using a case study driven by ridesharing data from Didi Chuxing.   In particular, we use data from Didi Chuxing \cite{fang2017prices} to reverse engineer the utility functions in the model. This yields the following form:
\begin{equation}\label{eq:f}
f(x)= \frac{1}{\gamma}(x-x\log x),
\end{equation}
where $\gamma=0.832$.
In our case study we compute the revenue and the optimal loyalty program under different price and heterogeneity settings, based on the user utility in (\ref{eq:f}).
Note that $f(x)$ is strictly concave and increasing in $[0,1]$, with a derivative of $0$ at $x=1$.

\subsection{Monopolistic market}\label{subsection:homo-market}

We start by considering a monopolistic market and provide the following results.

\textbf{How does the supplier base payment $p$ impact revenue?} We first demonstrate that the optimal pricing strategy for a firm is to set the base payment $p=0$ and extract revenue via the loyalty program.

Specifically, we fix the renter charge $q=12$ and consider a homogeneous market, which adopt the optimal linear loyalty program derived in Theorem \ref{thm:lp_mono}. Then, we  vary the value $p$ and compute the revenue.
%
From Figure \ref{fig:monopoly}, we see that the platform's revenue is decreasing in the supplier price $p$. Note that this is not obvious, since a larger $p$  facilitates sharing and may lead to a larger total supply. Nonetheless, the results in Figure \ref{fig:monopoly}  illustrate Corollary \ref{thm:opt_mono} and shows that the platform should choose $p=0$ for base payment and rely on the loyalty program for extracting optimal revenue.


\textbf{How effective is HB?} We now
 investigate the effectiveness of the HB program in a heterogeneous market, and show that the revenue under HB is close to the optimal.

 Specifically, we choose the self-usage utility of owner $i$ to be:
\begin{equation}
	f_i(x)=\bigg(n-i+1\bigg)\cdot f(x),\,\,\,i=1,2,...,n.
\end{equation}

We set the base payment to $p=0$ (this is the optimal case), $q =12$, and vary the  number of owners $n$.
Figure \ref{fig:multi_revenue} shows the revenue and platform's total subsidy under the Hyperbolic bonus program ($R_{HB}$).
It can be observed that HB achieves good revenue performance since  $R_{HB}$ is close to the optimal revenue $R_{OPT}$.
Further, we can see that the total subsidy paid by platform is a constant beyond the opportunity cost,  i.e., the extra payment is independent of $n$.
This validates Theorem \ref{thm:hb}.
Note that when $n=1$, the two curves coincide, as the market is homogeneous and the single threshold linear loyalty program is optimal.



\begin{figure*}[t]
	\begin{minipage}[t]{0.33\linewidth}
		\centering
		\includegraphics[scale=0.23]{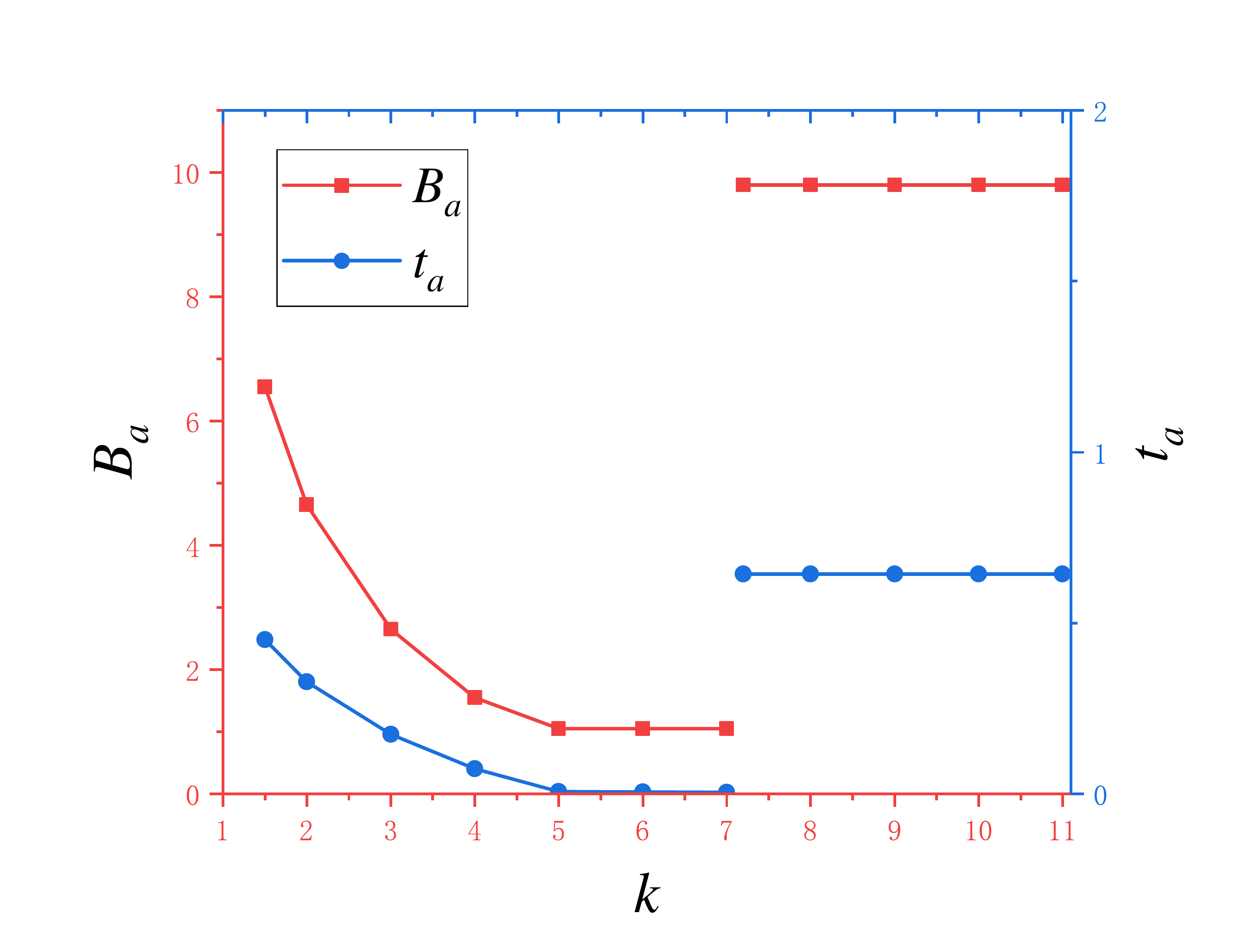}
		\vspace{-0.2in}
		\captionsetup{width=.9\linewidth}
		\caption{Platform $a$'s optimal linear loyalty program is to attract both owners when $k$ (user heterogeneity) is small, and will focus only on owner $2$ when $k$ is large ($p_a=10$, $p_b=11$).}
		\label{fig:lp_k}
	\end{minipage}%
	\begin{minipage}[t]{0.33\linewidth}
		\centering
		\includegraphics[scale=0.23]{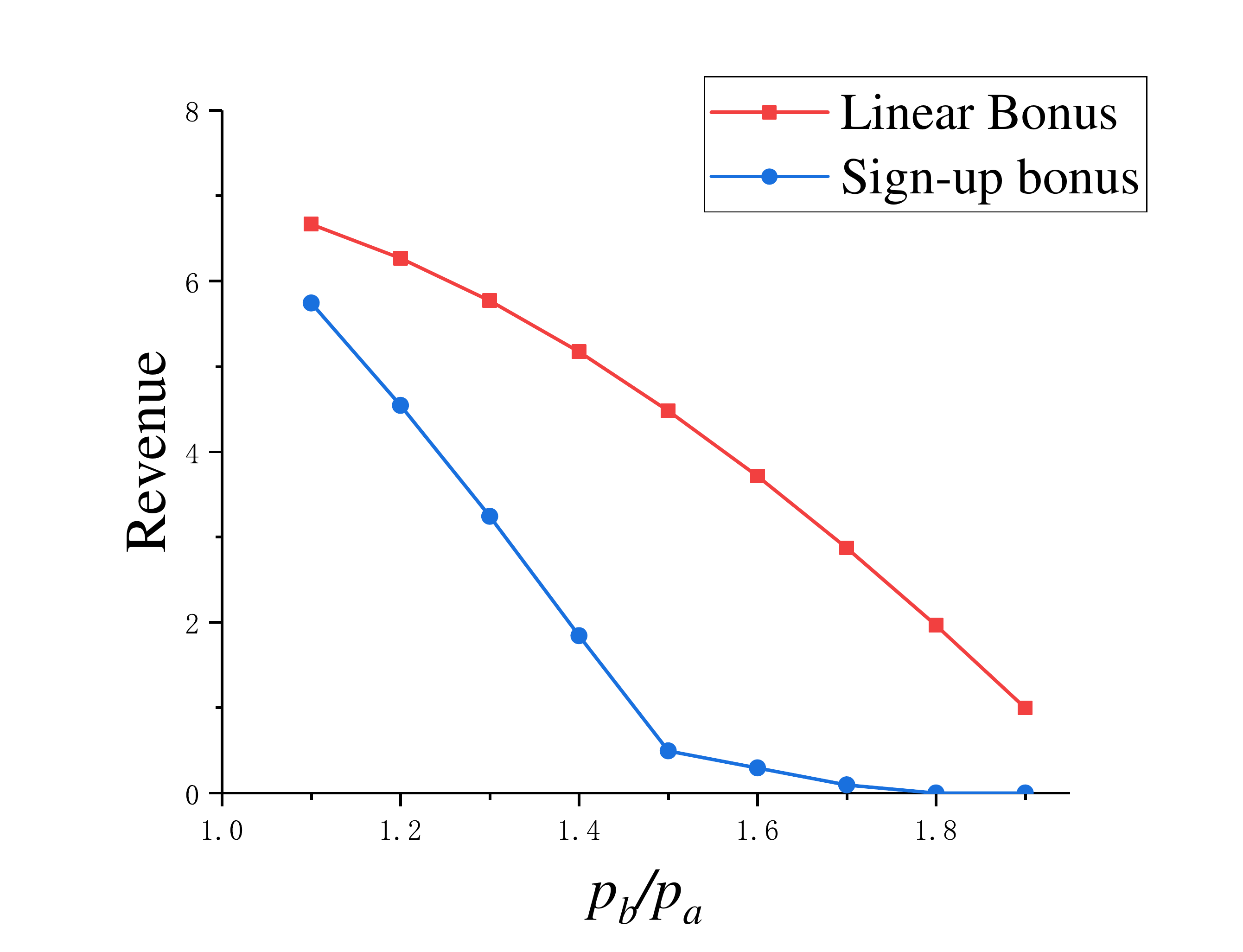}
		\vspace{-0.2in}
		\captionsetup{width=.9\linewidth}
		\caption{Platform $a$'s revenue. When only platform $a$ adopts subsidy program, the optimal linear loyalty program always brings a higher revenue than sign-up bonus ($p_a=10$).}
		\label{fig:rev_sb_lp}
	\end{minipage}
	\begin{minipage}[t]{0.33\linewidth}
		\centering
		\includegraphics[scale=0.23]{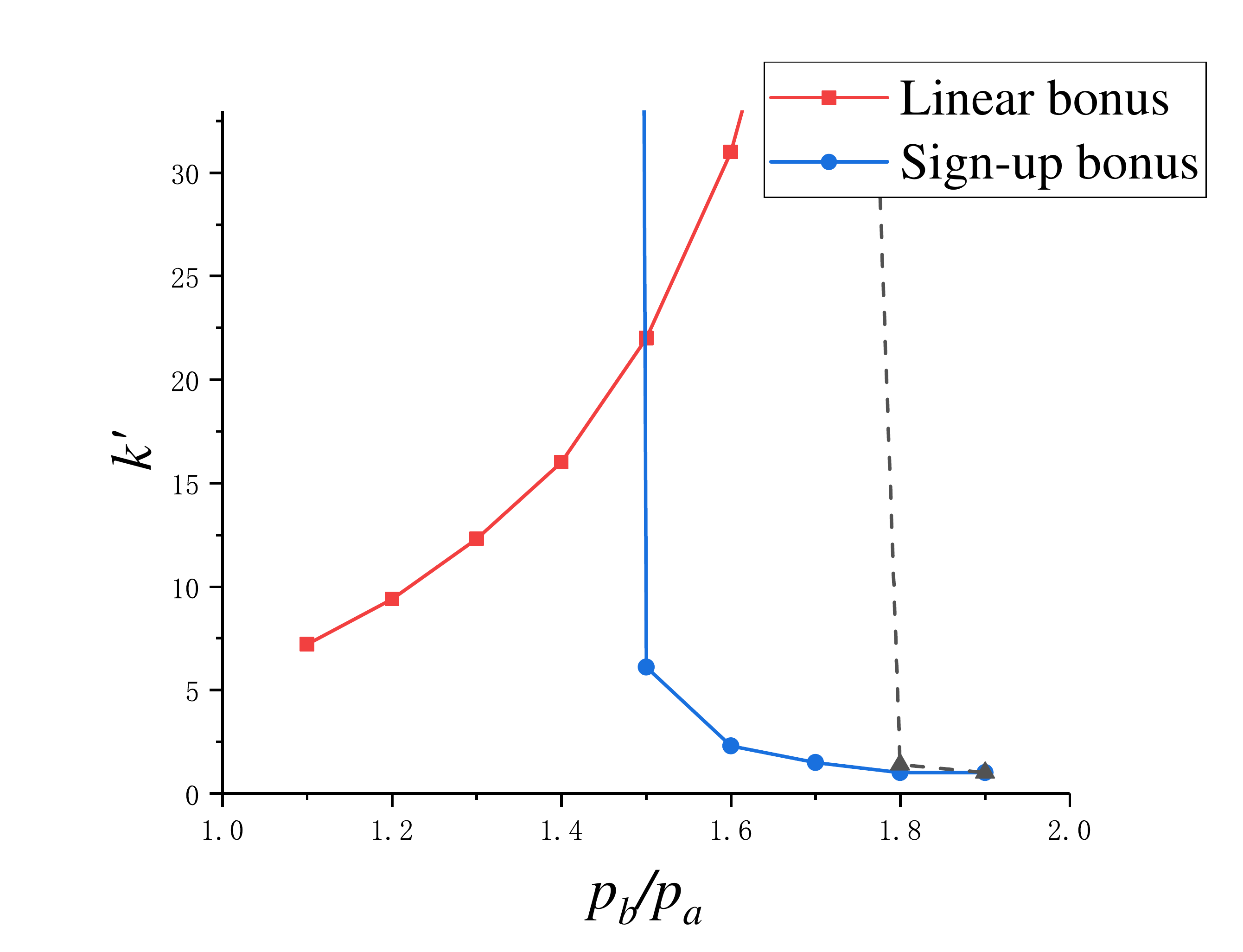}
		\vspace{-0.2in}
		\captionsetup{width=.9\linewidth}
		\caption{The critical $k'$ value beyond which platform $a$ switches from attracting both owners to only attracting one owner ($p_a=10$).}
		\label{fig:critical_k}
	\end{minipage}
	\vspace{-0.1in}
\end{figure*}

\subsection{Competitive market}\label{subsectiobn:compete}
Next, we consider loyalty program under competition.  
In this case, we capture the user heterogeneity with a parameter $k$ by
choosing $f_2(x)=f(x)$ and $f_1(x)=k\cdot f(x)$, i.e., $k$ being the difference between the two classes of owners.
We set $p_a=10$ in the experiments,  and let $\beta_a=\beta_b=1$.

\textbf{How does competition impact the optimal sign-up bonus?}
Consider platform $a$ adopts sign-up bonus and platform $b$ does not adopt loyalty program.
We first compute the optimal bonus  $B_a$ under different competition profiles $(p_a, p_b)$ with fixed $p_a$, and show that user heterogeneity plays a role in deciding market partition.

%
%
Figure \ref{fig:sb_p2} shows the result for $k=6$, where the optimal sign-up bonus is computed for three regimes.
\begin{enumerate}[(i)]
	\item The first regime is when $\frac{p_b}{p_a}<1.5$. In this regime, the sign-up bonus is rapidly increasing in $p_b$, and platform $a$'s best strategy is to attract both owners. 
	\item In the second regime, where the competition is tougher for platform $a$ (a larger $p_b$), platform $a$'s optimal strategy is to only attract owner $1$ by reducing revenue.  Thus, the bonus increases in $p_b$.
	\item In the last regime, $p_b$ is large enough such that $\frac{p_b}{p_a}>1.8$. In this case, platform $a$ cannot stop owners from leaving for platform $b$. As a result, it achieves zero revenue.
\end{enumerate}


\textbf{How does competition impact the form of the optimal linear loyalty program?}
We now study the form of the optimal linear loyalty program for platform $a$ under different $k$ values, while platform $b$ does not adopt loyalty program.
Platform $a$ will only attract one owner when user heterogeneity is extensive.
Figure \ref{fig:lp_k} shows the case for $p_a=10$ and $p_b=11$.
There are two regimes for the optimal linear loyalty program. 
\begin{enumerate}[(i)]
	\item If $k<7.2$,
	platform $a$'s best strategy is to adopt an optimal bonus $B^*$  to attract both owners and obtain higher revenue (exact form can be found in the proof of Theorem \ref{thm:lp}).
	\item If $k\geq 7.2$, i.e., when users are highly heterogeneous, platform $a$'s optimal strategy is to keep $B_a=q_a-p_a$,  and to target only owner $2$.
\end{enumerate}
The existence of transition point $k'=7.2$ shows the watershed in platform's subsidy policy, i.e.,  when users are highly differentiated ($k>k'$), it will be too expensive to attract both owners. In that case, platform $a$ chooses to stay with the more profitable owner $2$ and substantially increases the minimum sharing requirement to block owner $1$ from entering the platform.

\textbf{Do linear loyalty programs provide a competitive advantage?}
We now show that linear loyalty programs always lead to a higher revenue than sign-up bonuses in the competitive setting.

Figure \ref{fig:rev_sb_lp} shows the results for $k=6$ and different competition profiles $(p_a, p_b)$ when only platform $a$ adopts loyalty program. It can be observed that  linear loyalty program always leads to a higher revenue than sign-up bonus.
The reason is that linear bonuses can substantially boost supply, since  owners earn more bonus only if they share more. On the other hand, sign-up bonus does not increase the sharing amount.

\textbf{How does heterogeneity impact competition in providers?}

We now move to studying the impact of supplier heterogeneity on the optimal subsidizing policy, by finding the critical heterogeneity level beyond which platform $a$ switches to attract only one product owner.
%
%
In Figure \ref{fig:critical_k}, we plot the critical point $k'$ under different competition profiles $(p_a,p_b)$ when platform $a$ adopts linear loyalty program and sign-up bonus, respectively.


The blue line shows the critical $k'$ values of sign-up bonus. We see that $k'$ is very high ($>1000$) before $p_b/p_a<1.4$, and drops quickly to $k'=6.1$ when $p_b/p_a=1.5$. Then, it continues to decrease as $p_b$ increases.
This means that sign-up bonus can win both owners only when competitor's price $p_b$ or user heterogeneity $k$ is small.
The intuition behind this behavior is as follows. When $p_b$ is high, owner $2$'s sharing level at platform $b$ will be high. Thus, if platform $a$ wants to attract  owner $2$ (the one with a smaller marginal self-usage utility), it has to compensate owner $2$'s loss due to switching by giving a bonus that is larger than this loss. However, since $p_a$ is not large and the sign-up bonus itself does not facilitate sharing, attracting owner $2$ is not beneficial for platform $a$. Hence, it will try to instead only focus on attracting owner $1$ once $p_b/p_a$ is large.
The dash line shows the  value of $k$, below which using sign-up bonus can still gain positive revenue.

%
%
%

The red line shows the critical $k'$ values for linear loyalty bonuses.
Interestingly, $k'$ is increasing in $p_b$. This shows that linear loyalty program is more likely to win both owners, even when both $p_b/p_a$ and heterogeneity increase.
The reason that the critical point $k'$ is increasing  is as follows. First, the bonus for platform $a$ to attract owner $2$ is always  $B_a=q_a-p_a$. Hence, if $p_b$ is high, the only way platform $a$ can attract owner $2$ is by lowering the threshold value $t_a$. Moreover, the value $t_a$ decreases with $p_b$. Thus, if $p_b$ is high, platform already uses a small $t_a$. In this case, it only has to pay a smaller  additional bonus to also attract owner $1$ (the one with higher marginal self-usage utility), and by doing so, it also obtains owner $1$'s sharing, which brings additional benefit.
As a result, platform $a$ will stick to attracting both users until $k'$ is higher.
%
%
%
Note that this implies a higher robustness of the linear program, which is a desired feature  for platform strategies, especially at the early stage, when a company does not have enough data to accurately profile user behavior and heterogeneity.

\section{Conclusion}
In this paper, we consider the problem of optimal loyalty program (subsidy) design, which is central to the development of the sharing economy. We first characterize the optimal  loyalty program for a monopolistic market with homogeneous product owners, and show that it has a linear form that is close to loyalty programs in practice.
After that, we show that the multi-threshold loyalty program (MTLP) can achieve optimal revenue when the platform is facing heterogeneous suppliers.
Also, we introduce a hyperbolic bonus program (HB) for the heterogeneous market, and show that this program is highly cost-efficient and achieves  a good revenue performance.
Further, we consider loyalty programs in a duopoly market with heterogeneous suppliers, showing that the introduction of loyalty programs can help a weaker platform to win customers, and the more sophisticated ladder-like loyalty program can outperform the simple sign-up bonus.
Our results show that heterogeneity in users helps reduce the competition among platforms, since platforms tend to attract different groups of users in this case. We further validate our results with real transaction data from ride sharing platform Didi Chuxing.

\begin{acks}
The work of Zhixuan Fang and Longbo Huang is supported by the National Natural Science Foundation of China Grant 61672316 and 61303195, the Tsinghua Initiative Research Grant and the China Youth 1000-Talent Grant. The work of Adam Wierman is supported by NSF Grant AitF-1637598, CNS-1518941.
\end{acks}

\bibliographystyle{abbrv}
\bibliography{acmsmall-market-bibfile}


\newpage
\section{Appendix: Proofs} \label{sec:analysis}
This section provides proofs of the results stated and discussed in the previous sections.

\subsection{Proof of Theorem \ref{thm:lp_mono}} \label{subsec:lp_mono}
We first show that the linear loyalty program is optimal in a monopolistic market with homogeneous owners ($f_i(x)=f(x)$ for all $i\in \mathcal{O}$), given $p$ and $q$).
Since all owners are homogeneous, we can normalize the number of owners to be $1$ without loss of generality. In this case, we have $|\mathcal{O}|=1$ and thus the total supply $S=s$, where $s$ is the owner's sharing level. We then derive the optimal $B, t$.
The platform's revenue is the commission fee subtracting subsidies, as defined in (\ref{eq:revenue_loyal}):
\begin{equation}
R=(q-p) s - W(s).
\end{equation}

(i) Suppose there exists a general subsidy program $\tilde{B}(s)$ such that platform revenue is maximized.
We want to show that there exists a linear loyalty program with $(B^*,t)$ that can achieve the same revenue.
Denote the sharing level of owner under $\tilde{B}(s)$ to be $\tilde{s}$, let
\begin{equation*}
B^*=f'(1-\tilde{s})-p.
\end{equation*}
Since $\tilde{B}(s)$ maximizes $R$, the minimum subsidy is to compensate owners such that the total utility would be the same whether or not participating loyalty program $\tilde{B}(s)$.
Similarly, we can set $t$ such that
\begin{equation} \label{eq:loyalty_mono}
B^* \cdot (\tilde{s}-t)=\int_{1-\tilde{s}}^{1-s_0}(f'(x)-p) \ud x,
\end{equation}
where $s_0$ is the sharing level when there exists no loyalty program, which satisfies the marginal condition:
\begin{equation*}
f'(1-s_0)=p.
\end{equation*}
Since $\tilde{B}(s)$ maximizes revenue, we should at least have $\tilde{s}\geq s_0$.
Note that
\begin{align*}
\int_{1-\tilde{s}}^{1-s_0}(f'(x)-p)\ud x < & (f'(1-\tilde{s})-p)(\tilde{s}-s_0)\\
=& B^* (\tilde{s}-s_0).
\end{align*}
Here we use the marginal condition $f'(1-\tilde{s})=p+B$.
Thus there must exists a $t\in[s_0,\tilde{s}]$ which satisfies (\ref{eq:loyalty_mono}).
In conclusion, given $(B^*, t)$ above, owner's sharing level and platform's revenue are the same as the optimal general subsidy $\tilde{B}(s)$.

(ii) Similar to analysis above, the linear loyalty program with parameter $B,t$ achieves maximum revenue when:
\begin{align}\label{eq:homo_smin}
B \cdot (s-t)=\int_{1-s}^{1-s_0}(f'(x)-p)\ud x.
\end{align}
Therefore platform's revenue is:
\begin{align*}
R &= (q-p)s-B(s-t) \nonumber\\
& = (q-p) s -  \int_{1-s}^{1-s_0}(f'(x)-p)\ud x
\end{align*}
consider first order condition,
\begin{align*}
\frac{dR}{ds}&=(q-p) - (f'(1-s)-p)\nonumber\\
&=(q-p)- B =0,
\end{align*}
which gives $B = q-p =\beta p$. Here we use the marginal utility condition that  $f'(1-s)=B+p$.
The bonus threshold $t$ is given by (\ref{eq:homo_smin}):
\begin{equation*}
t=s-\frac{f(1-s_0)-f(1-s)-p(s-s_0)}{q-p}.
\end{equation*}

\subsection{Proof of Corollary \ref{thm:opt_mono}}
By analysis in the proof of Theorem \ref{thm:lp_mono}, the maximum revenue of platform under price $p$ and loyalty program $(B,t)$ is:
\begin{align*}
R(p,B)& =(q-p)s -B(s-t)\nonumber\\
&=(q-p)s-\int_{1-s}^{1-s_0}(f'(x)-p)\ud x \nonumber\\
&=qs -ps_0 -\int_{1-s}^{1}f'(x)\ud x +\int_{1-s_0}^{1}f'(x)\ud x,
\end{align*}
where $s_0$ is the sharing level when there is no loyalty program.
Therefore it is platform's strategy to choose supply payment $p$ and bonus $B$ (which will uniquely determine $t$ by (\ref{eq:homo_smin})).
Under first order condition:
\begin{align*}
\frac{\partial R}{\partial B} &= q \frac{\partial s}{\partial B} -f'(1-s)\frac{\partial s}{\partial B} \nonumber\\
&=\frac{\partial s}{\partial B}(q-p-B)\nonumber\\
&=0,
\end{align*}
which gives optimal $B=q-p$. Note that $q$ is fixed and $s_0$ is independent of $B$ here.
Similarly,
\begin{align*}
\frac{\partial R}{\partial p} &= q \frac{\partial s}{\partial p} -s_0-p\frac{\partial s_0}{\partial p}-f'(1-s)\frac{\partial s}{\partial p}+f'(1-s_0)\frac{\partial s_0}{\partial p} \nonumber\\
&= (q-f'(1-s)) \frac{\partial s}{\partial p} -(p-f'(1-s_0))\frac{\partial s_0}{\partial p} -s_0 \nonumber\\
&= -s_0<0.
\end{align*}
The last equation is due to the fact that $p=f'(1-s_0)$, $B=q-p$ and $p+B=f'(1-s)$ derived above.
The result shows that revenue $R$ is decreasing in $p$, which means optimal supply payment should be $p=0$, and introduces loyalty program such that :
\begin{align*}
B&=q,\\
t&=s-\frac{f(1)-f(1-s)}{q}.
\end{align*}


%

\subsection{Proof of Proposition \ref{prop:non-decreasing}}
Given $n$ owners with self usage benefit $f_i(x)$, let $s_{i,0}$ denote the sharing level of $i$ under no loyalty program, i.e., $f'_i(1-s_{i,0})=p$.
Consider some owner $i$ with sharing level $s_i$ under loyalty bonus with non-decreasing $B(s)$, such that $s_i>s_{i,0}$, i.e, one who participates in loyalty program $B(s)$ and thus shares more than initial sharing $s_{i,0}$ (it will be trivial if none of owners participate).
We must have the bonus covers owner $i$'s opportunity cost:
\begin{equation}
\int_{0}^{s_i}\bigg(p+B(s)\bigg)\ud s \geq \int_{1-s_i}^{1}f_i(x)\ud x,
\end{equation}
and the marginal condition that $f_i(1-s_i)=p+B(s_i)$.

We  show that for any $j>i$, we must have owner $j$ also participates in the loyalty program.
Note that $p+B(s_i)=f'_i(1-s_i)>f'_j(1-s_i)$. If $p+B(s)>f'_j(x)$ for all  $x<1-s_i$, owner $j$ must participate in the program and share $s_j\geq s_i$.
Otherwise, there must exists some intersection point $x_j$ such that $f'(x_j)=p+B(1-x_j)$ in $x_j\in[0,1-s_i]$.

Let $\tilde{x_j}=\max\{x|f_j'(x)=p+B(1-x),x\in[0,1-s_i]\}$, we must have $p+B(s)>f'_j(x)$ for $x\in[\tilde{x_j},1-s_i]$.
Therefore, we have:
\begin{align*}
&\int_{0}^{1-\tilde{x_j}}\bigg(p+B(s)\bigg)\ud s \\
= & \int_{0}^{s_i}\bigg(p+B(s)\bigg)\ud s + \int_{s_i}^{1-\tilde{x_j}}\bigg(p+B(s)\bigg)\ud s\\
> & \int_{1-s_i}^{1}f'_i(x)\ud x +\int_{\tilde{x_j}}^{1-s_i }f'_j(x)\ud x.\\
\geq & \int_{\tilde{x_j}}^{1}f'_j(x)\ud x.
\end{align*}
In this case, owner $j$ will choose to share $1-\tilde{x_j}$ to earn more from the loyalty program.

\subsection{Proof of Theorem \ref{thm:multi_opt}}
We prove the theorem by showing that for any subsidy program with increasing marginal bonus, multi-threshold loyalty program can achieve the same revenue.

Suppose under the non-decreasing marginal bonus $B'(s)$, owner $i$'s sharing level is $s'_i$, we must have $s'_i\geq s'_{i,0}$, and $s'_n\geq s'_{n-1}\geq ...\geq s'_{1}$. By Proposition \ref{prop:non-decreasing}, assume owners $i\geq i_0$ joins the loyalty program.
If there is only one owner who joins the program, i.e., $i_0=n$, the proof is the same as in the Theorem \ref{thm:lp_mono}.
Otherwise, consider owner $i$ such that $i_0< i \leq n$.
There must exists some $t_i\in(s'_{i-1},s'_{i})$ such that:
\begin{align}\label{eq:b'_t}
B'(s'_i)(s'_i-t_i)+B'(s'_{i-1})(t_i-s'_{i-1})=\int_{s'_{i-1}}^{s'_i}B'(s)\ud s
\end{align}
For the case when  $B'(s'_i)=B'(s'_{i-1})$, we can choose any $t_i\in(s'_{i-1},s'_{i})$.
We also have $t_0\in(0,s'_{i_0})$ such that:
\begin{align}\label{eq:b'_t0}
B'(s'_{i_0})(s'_{i_0}-t_{i_0})=\int_{0}^{s'_{i_0}}B'(s)\ud s
\end{align}

Construct a multi-threshold loyalty program such that $B_i=B'(s_i)$, and threshold $t_i$ according to (\ref{eq:b'_t}) and (\ref{eq:b'_t0}), for all $i\geq i_0$.
In this case, each owner $i$ will share the same $s'_i$.
Note the utility for $i$ to share $s'_i$ stays the same in MTLP, compared to $B'(s)$.
The reason is that $i$ will only choose to share at intersection points of $f'_i(x)$ and $B(s)$, i.e.,  where $f'_i(1-s)=B(s)$, if any such $s$ exists.
For such point $s$ such that $s>s'_i$, and $s\in[t_k,t_{k+1})$ where $k>i$, we have:
\begin{align*}
&B_k(s-t_k)+\sum_{j=i+1}^{k-1}B_j(t_{j+1}-t_{j})+B_i(t_{i+1}-s'_{i})\\
\leq & \int_{s'_i}^{s}B'(s)\ud s < \int_{1-s}^{1-s'_i}\bigg(f'_i(x)-p\bigg)\ud x
\end{align*}
given $i$ choose $s'_i$ as optimal sharing under $B'(s)$.
The first inequality is due to the fact that $B_k(s-t_k)+B_{k-1}(t_k-s'_{k-1})\leq \int_{s'_{k-1}}^{s}B'(s)\ud s$ for $s\in[s'_{k-1},s'_{k})$.
This means $i$ has no motivation to over share under MTLP $B(s)$.
Similarly, if $i$ choose a smaller  sharing level $s$ such that $s\in[t_k,t_{k+1}]$ where $k<i$, we have:
\begin{align*}
&B_i(s'_i-t_i)+\sum_{j=k+1}^{i-1}B_j(t_{j+1}-t_{j})+B_k(t_{k+1}-s)\\
\geq &\int_{s}^{s'_i}B'(s)\ud s>\int_{1-s'_i}^{1-s}(f'_i(x)-p)\ud x,
\end{align*}
which also guarantees $i$ will not share less. The first inequality here is due to the fact that $B_{k+1}(s'_{k+1}-t_{k+1})+B_{k}(t_{k+1} - s)>\int_{s}^{s'_{k+1}}B'(s)\ud s$ for $s\in[s'_k,s'_{k+1})$.

Hence, the MTLP constructed here gives the same sharing level of each owner, and the revenue of platform is also the same:
\begin{align*}
R_{B'}=&\sum_{i=1}^{n}\bigg((q-p)s'_i-\int_{0}^{s'_i}B'(s)\ud s\bigg)\\
=&\sum_{i=1}^{n}\bigg((q-p)s'_i-B_i(s'_i-t_i)-\sum_{j=i_0}^{i-1}B_j(t_{j+1}-t_{j})\bigg)\\
=&R_{MTLP}.
\end{align*}
In conclusion, for any bonus with nondecreasing marginal $B'(s)$, we can construct a multi-threshold loyalty program to achieve the same revenue and sharing level, i.e., MTLP can achieve the optimal  platform revenue.

\subsection{Proof of Theorem \ref{thm:multi_necessary}}
Note that $s_{i,i-1}$ satisfies $f'_i(1-s_{i,i-1})=B_{i-1}$.
If $t_i>s_{i,i-1}$, since owner $i$'s optimal sharing is $s_i$, we must have the income of sharing $s_i$ instead of $s_{i,i-1}$ is higher than the lost of self usage benefit from $1-s_{i,i-1}$ to $1-s_i$:
\begin{align*}
B_i(s_i-t_i)+B_{i-1}(t_i-s_{i,i-1})\geq \int_{1-s_i}^{1-s_{i,i-1}}f'_i(x)\ud x,
\end{align*}
which is exactly (\ref{eq:multi_constraint}).
Otherwise if $t_i<s_{i,i-1}$, since $B_i>B_{i-1}$ and $B_i=f'_i(1-s_i)>f'_i(1-s)$ for $s<s_i$,  we always have:
\begin{align*}
B_i(s_i-t_i)\geq B_{i-1}(s_{i,i-1}-t_i)+\int_{1-s_i}^{1-s_{i,i-1}}f'_i(x)\ud,
\end{align*}
which also gives (\ref{eq:multi_constraint}).

\subsection{Proof of Theorem \ref{thm:hb}}
From previous analysis, the bonus should at least cover owner's opportunity cost, and $p=0$  gives the minimum opportunity cost of owners.
Note that by adding $B_{i-1}(s_{i,i-1}-t_{i-1})$ on both sides of (\ref{eq:multi_constraint}), we have:
\begin{align}\label{eq:multi_add_both_side}
&B_i(s_i-t_i)+B_{i-1}(t_{i}-t_{i-1})\nonumber\\
\geq&\int_{1-s_{i}}^{1-s_{i,i-1}}f_i'(x)\ud x+B_{i-1}(s_{i,i-1}-t_{i-1}), \,\,\,\forall i,
\end{align}
(i) According to (\ref{eq:rev_multi}), when a platform adopts multi-threshold loyalty program, its revenue is:
\begin{align*}
&R=q\sum_{i=1}^{n}s_i -\sum_{i=1}^{n}\bigg[B_{i}(s_{i}-t_{i})+\sum_{j=1}^{i-1}B_{j}(t_{j+1}-t_{j})\bigg] \\
\stackrel{(a)}{\leq}&q\sum_{i=1}^{n}s_i - \sum_{i=1}^{n}\bigg[\int_{1-s_{i}}^{1-s_{i,i-1}}f_i'(x)\ud x+B_{i-1}(s_{i,i-1}-t_{i-1})\\
&+\sum_{j=1}^{i-2}B_{j}(t_{j+1}-t_{j})\bigg]\\
=& q\sum_{i=1}^{n}s_i -\sum_{i=1}^{n}\int_{1-s_{i}}^{1-s_{i,i-1}}f_i'(x)\ud x-\sum_{i=1}^{n-1}B_i(s_{i+1,i}-t_{i})\\
&-\sum_{i=1}^{n-2}(n-i-1)B_i(t_{i+1}-t_{i})\\
\stackrel{(b)}{\leq}& q\sum_{i=1}^{n}s_i - \sum_{i=1}^{n}\int_{1-s_{i}}^{1-s_{i,i-1}}f_i'(x)\ud x -\sum_{i=1}^{n-1}B_i(s_{i+1,i}-t_{i})\\
&-\hspace{-0.05in} \sum_{i=1}^{n-2}(n-i-1)\bigg[\int_{1-s_{i+1}}^{1-s_{i+1,i}}\hspace{-0.08in} f_{i+1}'\ud x +B_i(s_{i+1,i}-t_{i})-B_{i+1}(s_{i+1}-t_{i+1})\bigg] \\
=& q\sum_{i=1}^{n}s_i - \sum_{i=1}^{n}\int_{1-s_{i}}^{1-s_{i,i-1}}f_i'(x)\ud x -\sum_{i=1}^{n-2}(n-i)B_i(s_{i+1,i}-t_{i}) \\
& -B_{n-1}(s_{n,n-1}-t_{n-1}) -\sum_{i=2}^{n-1}(n-i)\int_{1-s_{i}}^{1-s_{i,i-1}}f_i'(x)\ud x  \\
&  +\sum_{i=2}^{n-1}(n-i)B_i(s_i-t_{i})\\
=& q\sum_{i=1}^{n}s_i -\sum_{i=2}^{n-1}(n-i)\int_{1-s_{i}}^{1-s_{i,i-1}}f_i'(x)\ud x -\sum_{i=2}^{n-2}(n-i)B_i(s_{i+1,i}-s_{i})  \\
&-(n-1)B_1(s_{2,1}-t_{1})-B_{n-1}(s_{n,n-1}-s_{n-1})- \hspace{-0.05in} \sum_{i=1}^{n}\int_{1-s_{i}}^{1-s_{i,i-1}}\hspace{-0.05in}f_i'(x)\ud x\\
\stackrel{(c)}{\leq}& q\sum_{i=1}^{n}s_i -\sum_{i=2}^{n-1}(n-i)\int_{1-s_{i}}^{1-s_{i,i-1}}f_i'(x)\ud x - \sum_{i=1}^{n}\int_{1-s_{i}}^{1-s_{i,i-1}}f_i'(x)\ud x\\
& -\sum_{i=1}^{n-1}(n-i)B_i(s_{i+1,i}-s_{i})-(n-1)\int_{1-s_{1}}^{1}f'_1(x)\ud x\\
=& \sum_{i=1}^{n}[qs_i-(n-i+1)\int_{1-s_{i}}^{1-s_{i,i-1}}f_i'(x)\ud x] -\sum_{i=1}^{n-1}(n-i)B_i(s_{i+1,i}-s_{i}),
\end{align*}
where $(a)$ is to plug in (\ref{eq:multi_add_both_side}) to replace the term $B_{i}(s_{i}-t_{i})+B_{i-1}(t_{i}-t_{i-1})$, $(b)$ is to plug in (\ref{eq:multi_add_both_side}) to replace the term $(n-i-1)B_i(t_{i+1}-t_{i})$, and $(c)$ is to plug in  (\ref{eq:multi_add_both_side}) and use $(n-1)B_1(s_{2,1}-t_{1})=(n-1)B_1(s_{2,1}-s_{1})+(n-1)B_1(s_1-t_1)$.

Note that given (\ref{eq:multi_constraint_eq}), above $(a)$, $(b)$, $(c)$ become strict equality.

%

(ii)Hence, given sharing level of $(s_1,s_2,...,s_n)$, the optimal revenue is upper bounded by:
\begin{equation}\label{eq:revenue_upper}
R_{upper}= \sum_{i=1}^{n}(q s_i -\int_{1-s_i}^{1}f'_i(x)\ud x )
\end{equation}
Now we derive the gap between $R_{HB}$ and $R_{Upper}$, to show that HB is almost only covering owners' opportunity cost.
\begin{align*}
&R_{upper}-R_{HB}\\
=& \sum_{i=1}^{n}(n-i+1)\int_{1-s_{i}}^{1-s_{i,i-1}}f_i'(x)\ud x + \sum_{i=1}^{n-1}(n-i)B_i(s_{i+1,i}-s_{i})\\
&-\sum_{i=1}^{n}\bigg[\int_{1-s_{i}}^{1-s_{i,i-1}}f'_i(x)\ud x+\int_{1-s_{i,i-1}}^{1}f'_i(x)\ud x\bigg]\\
=&\sum_{i=1}^{n-1}(n-i)\bigg[\int_{1-s_{i}}^{1-s_{i,i-1}} \hspace{-0.1in} f_i'(x)\ud x + B_i(s_{i+1,i}-s_{i})\bigg]-\sum_{i=1}^{n}\int_{1-s_{i,i-1}}^{1} \hspace{-0.1in} f'_i(x)\ud x\\
=&\sum_{i=1}^{n-1}(n-i)\bigg[\int_{1-s_{i}}^{1} \hspace{-0.05in} f_i'(x)\ud x -\int_{1-s_{i,i-1}}^{1} \hspace{-0.05in} f'_i(x)\ud x + B_i(s_{i+1,i}-s_{i})\bigg]\\
&-\sum_{i=1}^{n}\int_{1-s_{i,i-1}}^{1}f'_i(x)\ud x\\
=&\sum_{i=1}^{n-1}(n-i)\bigg[\int_{1-s_{i+1,i}}^{1} \hspace{-0.08in} \min\{B_i,f'_i(x)\}\ud x - \int_{1-s_{i+1,i}}^{1} \hspace{-0.05in} f'_{i+1}(x)\ud x\bigg]\\
&- n \int_{1-s_{1,0}}^{1} \hspace{-0.05in} f'_{1}(x)\ud x+\sum_{i=1}^{n}\int_{1-s_{i,i-1}}^{1} \hspace{-0.05in} f'_i(x)\ud x-\sum_{i=1}^{n}\int_{1-s_{i,i-1}}^{1} \hspace{-0.05in} f'_i(x)\ud x\\
\end{align*}
The last equation is because  $B_i=f'_i(1-s_i)\geq f'_i(x)$ for $x\in[1-s_{i},1]$, and $B_i\leq f'_i(x)$ for $x\in[1-s_{i+1,i},1-s_{i}]$. Thus,
\begin{align*}
&\sum_{i=1}^{n-1}(n-i)\bigg[\int_{1-s_{i}}^{1}  f_i'(x)\ud x+ B_i(s_{i+1,i}-s_{i})\bigg]\\
=&\sum_{i=1}^{n-1}(n-i)\bigg[\int_{1-s_{i}}^{1}  f_i'(x)\ud x+ \int_{1-s_{i+1,i}}^{1-s_{i}} B_i\ud x\bigg]\\
=&\sum_{i=1}^{n-1}(n-i)\int_{1-s_{i+1,i}}^{1} \hspace{-0.08in} \min\{B_i,f'_i(x)\}\ud x.
\end{align*}
Continue to finish the proof, we have:
\begin{align*}
&R_{upper}-R_{HB}\\
\stackrel{(a)}{=}& \sum_{i=1}^{n-1}(n-i)\bigg[\int_{1-s_{i+1,i}}^{1} \hspace{-0.1in} \min\{\frac{q}{n-i+1},f'_i(x)\}\ud x - \int_{1-s_{i+1,i}}^{1} \hspace{-0.05in} f'_{i+1}(x)\ud x\bigg]\\
= & \sum_{i=1}^{n-1}\bigg[\int_{1-s_{i+1,i}}^{1} \hspace{-0.15in} \min\{\frac{q(n-i)}{n-i+1},(n-i)f'_i(x)\}\ud x - \int_{1-s_{i+1,i}}^{1} \hspace{-0.15in} (n-i)f'_{i+1}(x)\ud x\bigg]\\
\stackrel{(b)}{<}&\sum_{i=1}^{n-1}\bigg[\int_{1-s_{i+1}}^{1} \hspace{-0.16in} \min\{\frac{q(n-i)}{n-i+1},(n-i)f'_i(x)\}\ud x - \hspace{-0.03in} \int_{1-s_{i+1}}^{1} \hspace{-0.1in} (n-i-1)f'_{i+1}(x)\ud x\bigg]\\
<&\sum_{i=1}^{n-1}\bigg[\int_{1-s_{i+1}}^{1-s_{i}}q\frac{(n-i)}{n-i+1}\ud x + \int_{1-s_{i}}^{1}(n-i)f'_i(x)\ud x\\
&-\int_{1-s_{i+1}}^{1}(n-i-1)f'_{i+1}(x)\ud x\bigg]\\
=&\sum_{i=1}^{n-1}\bigg[\int_{1-s_{i+1}}^{1-s_{i}}q\frac{(n-i)}{n-i+1}\ud x\bigg] + (n-1) \int_{1-s_1}^{1}f'_{1}(x)\ud x\\
<&\sum_{i=1}^{n-1}q\frac{(n-i)}{n-i+1}(s_{i+1}-s_{i}) + q\frac{n-1}{n} s_{1}\\
=&\sum_{i=2}^{n-1}q s_i\bigg(\frac{1}{n-i+1}-\frac{1}{n-i+2}\bigg)-q\frac{n-1}{n}s_1+\frac{q}{2}s_n+ q\frac{n-1}{n} s_{1}\\
<&q\sum_{i=2}^{n-1} \bigg[\frac{s_{i+1}}{n-(i-1)}-\frac{s_i}{n-(i-2)}\bigg]+\frac{q}{2}s_n\\
<&q(s_{n}-\frac{s_2}{n})<q.
\end{align*}
Here $(a)$ is because of the fact that $s_{1,0}=0$ gives $\int_{1-s_{1,0}}^{1} f'_{1}(x)\ud x=0$, and we also plug in the Hyperbolic pricing in (\ref{eq:multi_bonus}).
Inequality $(b)$ is because $q\frac{n-i}{n-i+1}>q\frac{n-i-1}{n-i}=(n-i-1)f'_{i+1}(1-s_{i+1})>(n-i-1)f'_{i+1}(x)$ for $x\in[1-s_{i+1},1-s_{i+1,i}]$.

\subsection{Proof of Theorem \ref{thm:sb}}\label{subsec:sb}

Since $p_a<p_b$, platform $a$ is less attractive to product owners.
That means, given owner's utility in (\ref{eq:utility}), platform $a$ will have no supply and hence zero revenue. Thus, it is necessary for platform $a$ to introduce subsidy program.
Here recall that we consider \emph{exclusive} sign-up bonus, i.e., if  a product owner shares at platform $a$, he receives a one-time reward $B_a>0$ and does not share at platform $b$.

We now derive the optimal sign-up bonus for platform $a$.
Denote $x_{ij}$, $s_{ij}$ as the self-usage and sharing level of owner $i$  on platform $j$.
%
Since no owner will share simultaneously on two platforms, the utilities of  owner $i$ sharing on platform $a, b$, respectively, are given by:
\begin{align*}
U_{ia}=&f_i(x_{ia})+p_a s_{ia}+B_a\\
U_{ib}=&f_i(x_{ib})+p_b s_{ib}
\end{align*}
Note that if owner $i$ shares on platform $j$, then the following condition holds,
\begin{equation}
f_i'(x_{ij})=p_j, \label{eq:margin-p}
\end{equation}
where $f_i'(x)$ is the derivative of $f_i(x)$. Intuitively, this condition means that  the marginal income from self-usage and sharing should be equal.
Otherwise, one can allocate more resources towards the option with higher marginal income to gain higher utility.

If platform $a$ wants to attract owner $1$, its bonus should be high enough such that owner $1$ will decide to leave platform $b$, even though it offers a higher  price. Thus,
\begin{align*}
U_{1a}-U_{1b}&=(f_1(x_{1a})+p_a s_{1a}+B_a)-(f_1(x_{1b})+p_b s_{1b})\\
&=\int_{1-s_{1b}}^{1-s_{1a}}f_1'(x)\ud x+p_a s_{1a}-p_b s_{1b}+B_a\\
&=\int_{1-s_{1b}}^{1-s_{1a}}(f_1'(x)-p_a)\ud x-(p_b-p_a)s_{1b}+B_a\\
&\geq 0.
\end{align*}
Therefore, when bonus $B_a$ satisfies the following inequality,
\begin{equation}\label{eq:sb_b1_low}
B_a\geq (p_b-p_a)s_{1b}-\int_{1-s_{1b}}^{1-s_{1a}}(f_1'(x)-p_a)\ud x,
\end{equation}
owner $1$ will share on platform $a$. In this case, the revenue of platform $a$ is
\begin{align*}
R_{a}&=(q_a-p_a) s_{1a}-B_a\\
&\leq (q_a-p_a) s_{1a}-(p_b-p_a)s_{1b}+\int_{1-s_{1b}}^{1-s_{1a}}(f_1'(x)-p_a)\ud x\\
&\leq q_a s_{1a}-p_b s_{1b} +\int_{1-s_{1b}}^{1-s_{1a}}f_1'(x)\ud x.
\end{align*}
As a result, if platform $a$'s sign-up bonus is $B_a=(p_b-p_a)s_{1b}-\int_{1-s_{1b}}^{1-s_{1a}}(f_1'(x)-p_a)\ud x$, its revenue is maximized:
\begin{eqnarray}
R_{a}^{\mathbbm{1}}=q_a s_{1a}-p_b s_{1b} +\int_{1-s_{1b}}^{1-s_{1a}}f_1'(x)\ud  x > 0.
\end{eqnarray}
Here $R_{a}^{\mathbbm{1}}$ denotes the revenue of platform $a$ with only owner $1$ on it.
Plug in the fact that $\int_{1-s_{1b}}^{1-s_{1a}}f_1'(x)\ud  x>f_1'(1-s_{1a})(s_{1b}-s_{1a})$, we have the inequality condition in Theorem \ref{thm:sb} to guarantee a positive revenue.

Similar to (\ref{eq:sb_b1_low}), if platform $a$ wants to attract owner $2$, it needs:
\begin{equation}\label{eq:sb_b1_high}
B_a\geq (p_b-p_a)s_{2b}-\int_{1-s_{2b}}^{1-s_{2a}}(f_2'(x)-p_a)\ud x
\end{equation}
Note that the constraint (\ref{eq:sb_b1_high}) is tighter than (\ref{eq:sb_b1_low}), i.e., if (\ref{eq:sb_b1_high}) holds, (\ref{eq:sb_b1_low}) is always satisfied. This is because
\begin{align*}
&[(p_b-p_a)s_{2b}-\int_{1-s_{2b}}^{1-s_{2a}}(f_2'(x)-p_a)\ud x]\\
& -[(p_b-p_a)s_{1b}-\int_{1-s_{1b}}^{1-s_{1a}}(f_1'(x)-p_a)\ud x]\\
=&\int_{1-s_{2b}}^{1}(p_b-\max\{p_a,f_2'(x)\})\ud x-\int_{1-s_{1b}}^{1}(p_b-\max\{p_a,f_1'(x)\})\ud x\\
=&\int_{1-s_{2b}}^{1-s_{1b}}(p_b-\max\{p_a,f_2'(x)\})\ud x\\
&+\int_{1-s_{1b}}^{1}(\max\{p_a,f_1'(x)\}-\max\{p_a,f_2'(x)\})\ud x\\
>&0
\end{align*}
Therefore, if bonus $B_a$ is high enough such that owner $2$ only shares on platform $a$, owner $1$ will also choose to share on platform $a$.
Hence, the revenue of winning both owners will be:
\begin{align*}
R_{a}^{both}&=(q_a-p_a)(s_{2a}+s_{1a})-2B_a\\
&\leq (q_a+p_a)s_{2a}+(q_a-p_a) s_{1a}-2p_b s_{2b}+2\int_{1-s_{2b}}^{1-s_{2a}}f_2'(x)\ud x.
\end{align*}
In conclusion, given the condition in Theorem \ref{thm:sb}, platform $a$ can set its bonus high enough such that  (\ref{eq:sb_b1_low}) is satisfied to obtain a positive revenue.
%

\subsection{Proof of Theorem \ref{thm:lp}}\label{subsec:loyalty}
We consider the asymmetric case here, i.e.,  $p_a<p_b$. That means, platform $a$ has a lower supplier price.
%
%
In order to survive in the competition, platform $a$ can adopt the linear loyalty program to boost supply. Recall  the owner utility given in (\ref{eq:utility-lp}) and (\ref{eq:utility}), i.e.,
\begin{equation}
U_{i}=f_i(x_i)+p_a s_{ia}+p_b s_{ib}+B_a(s_{ia}-t_a)^+ \label{eq:linear bonus}
\end{equation}
where $B_a(s_{ia}-t_a)^+=\max(s_{ia}-t_a,0)$.

Note that the reasonable bonus should be chosen such that $B_a+p_a>p_b$. Otherwise, no owner will go to platform $a$ and the loyalty program simply does not work.
We first show that this linear bonus program (\ref{eq:linear bonus}) will lead to loyalty decisions.
\begin{theorem}\label{thm:exclusive}
	If platform $a$ adopts linear loyalty program and platform $b$ adopts no loyalty program, we have $s_{ia}\cdot s_{ib}=0$ for all owner $i$. That is, no owner will  share on two platforms simultaneously.
\end{theorem}

Next we show that platform $a$ can have positive revenue with the introduction of loyalty program.
Recall that since $f_i(x)$ is increasing in $[0,1]$, we always have $x_i+s_{ia}+s_{ib}=1$ for all owner $i$.  Our proof proceeds in four steps.

\textbf{Step (i): Derive the condition under which platform $a$ attracts owner $2$.} Since owners will only share on one platform, we denote  owner $2$'s  utility to be $U_{2j}$  when he chooses to share on platform $j$.
We denote $s_{2j}$  as owner $2$'s sharing level on platform $j$, and $x_{2j}$ as the self-usage level when owner $2$ shares on platform $j$.

Similar to (\ref{eq:margin-p}), we have
\begin{equation}\label{eq:user1_margin}
f_i'(x_i)=f_i'(1-s_{2a})=p_a+B_a,
\end{equation}
where $f_i'(x)$ is the derivative of $f_i(x)$.
Note that the reason for owner $2$ to come to platform $a$ must be the extra bonus. Thus, we have $t_a<s_{2a}$.
Now the utilities of $U_{2a}$ and  $U_{2b}$ are given by:
\begin{align*}
&U_{2a}=f_2(x_{2a})+p_a s_{2a}+B_a(s_{2a}-t_a)\\
&U_{2b}=f_2(x_{2b})+p_b s_{2b}
\end{align*}
Given $p_a+B_a>p_b$ and that $f_2(x)$ is strictly concave and increasing in $[0,1]$, we must have $x_{2a}<x_{2b}$. Due to $s_{2j}=1-{x_{2j}}$, we also have $s_{2a}>s_{2b}$.
Hence, we can rewrite $U_{2b}$ in the following form:
\begin{align*}
U_{2b}&=f_2(x_{2b})+p_b s_{2b}\\
&=f_2(x_{2a})+\int_{1-s_{2a}}^{1-s_{2b}}f_2'(x)\ud x +p_b s_{2b}\\
&=f_2(x_{2a})+\int_{1-s_{2a}}^{1-s_{2b}}(f_2'(x)-p_b)\ud x+p_b s_{2a}
\end{align*}
Thus, we need
\begin{align*}
U_{2a}-U_{2b}&=(p_a+B_a-p_b)s_{2a}-\int_{1-s_{2a}}^{1-s_{2b}}(f_2'(x)-p_b)\ud x-B_a t_a\\
&\geq 0
\end{align*}
which gives a condition:
\begin{equation}\label{eq:bs_bound}
B_a t_a\leq (p_a+B_a-p_b)s_{2a}-\int_{1-s_{2a}}^{1-s_{2b}}(f_2'(x)-p_b)\ud x
\end{equation}
Therefore, to attract owner $2$,   $t_a$ is upper bounded by (\ref{eq:bs_bound}) given platform $a$'s  bonus $B_a$.
We verify the right hand side of (\ref{eq:bs_bound}) to be always positive in the following:
\begin{align}\label{eq:verify_pos}
&(p_a+B_a-p_b)s_{2a}-\int_{1-s_{2a}}^{1-s_{2b}}(f_2'(x)-p_b)\ud x\nonumber\\
\geq & (p_a+B_a-p_b)s_{2a}-\int_{1-s_{2a}}^{1-s_{2b}}(p_a+B_a-p_b)\ud x\nonumber\\
=&(p_a+B_a-p_b)s_{2b}\nonumber\\
>&0.
\end{align}

Hence it is always possible to find a feasible $(B_a, t_a)$ pair.

\textbf{Step (ii): Derive the condition under which owner $1$ comes to platform $a$.}
Suppose platform $a$ has owner $1$ on the platform. In this case, the utilities of owner $1$ on each platform are given by:
\begin{align*}
U_{1a}&=f_1(x_{1a})+p_a s_{1a} +B_a(s_{1a}-t_a)\\
U_{1b}&=f_1(x_{1b})+p_b s_{1b}.
\end{align*}
Similar to (\ref{eq:bs_bound}),  we must have  owner $1$ obtaining a  higher utility on platform $a$ than on platform $b$, i.e.,
\begin{equation}\label{eq:bs_bound21}
B_a t_a\leq (p_a+B_a-p_b)s_{1a}-\int_{1-s_{1a}}^{1-s_{1b}}(f_1'(x)-p_b)\ud x.
\end{equation}
Similar to (\ref{eq:verify_pos}), the right hand side of (\ref{eq:bs_bound21}) is always positive.
Note that the bound for $B_a t_a$ in (\ref{eq:bs_bound21}) is tighter than that in (\ref{eq:bs_bound}) because:
\begin{align*}
&[(p_a+B_a-p_b)s_{2a}-\int_{1-s_{2a}}^{1-s_{2b}}(f_2'(x)-p_b)\ud x]\\
&-[(p_a+B_a-p_b)s_{1a}-\int_{1-s_{1a}}^{1-s_{1b}}(f_2'(x)-p_b)\ud x]\\
=&[\int_{p_b}^{p_a+B_a}s_{2a}\ud y-\int_{p_b}^{p_a+B_a}(f_2'^{-1}(y)-(1-s_{2a}))\ud y]\\
&-[\int_{p_b}^{p_a+B_a}s_{1a}\ud y-\int_{p_b}^{p_a+B_a}(f_1'^{-1}(y)-(1-s_{1a}))\ud y]\\
=&\int_{p_b}^{p_a+B_a}(f_1'^{-1}(y)-f_2'^{-1}(y))\ud y\\
>&0.
\end{align*}
Consequently, whenever platform $a$ has owner $1$, it also has owner $2$. Hence, (\ref{eq:bs_bound21}) is the condition for owner  $1$  to join platform $a$.

\textbf{Step (iii):
	Derive the optimal strategy for platform $a$ to maximize revenue, given that only owner $2$ is on platform $a$}.
In this case, we consider the objective of platform is to maximize the transaction volume subtracting bonus expenditure, i.e.,
\begin{equation}\label{eq:revenue_linear}
R_{a}=(q_a-p_a) s_{2a}-B_a(s_{2a}-t_a).
\end{equation}
Since the term $B_a s_{2a}$ is upper bounded in (\ref{eq:bs_bound}), the revenue is also upper bounded:
\begin{align}\label{eq:revenue_linear_s11}
R_{a}&=(q_a-p_a-B_a)s_{2a}+B_a t_a\nonumber\\
&\leq (q_a-p_a-B_a)s_{2a}+(p_a+B_a-p_b)s_{2a}-\int_{1-s_{2a}}^{1-s_{2b}}(f_2'(x)-p_b)\ud x\nonumber\\
&=(q_a-p_b)s_{2a}-\int_{1-s_{2a}}^{1-s_{2b}}(f_2'(x)-p_b)\ud x.
\end{align}
The revenue upper bound is achievable when
\begin{equation}\label{eq:s_min_one}
t_a=t_a^\mathbbm{2}=\frac{1}{B_a}[(p_a+B_a-p_b)s_{2a}-\int_{1-s_{2a}}^{1-s_{2b}}(f_2'(x)-p_b)\ud x].
\end{equation}
In order to maximize revenue, platform $a$ will choose $B_a$ and $t_a$ such that:
\begin{equation}
\frac{\ud R_{a}}{\ud s_{2a}}=0.
\end{equation}
Plug in (\ref{eq:revenue_linear_s11}), we have that:
\begin{align*}
\frac{\ud R_{a}}{\ud s_{2a}}&=q_a-p_b-f_2'(1-s_{2a})+p_b\\
&=q_a-f_2'(1-s_{2a})\\
&=q_a-(B_a^*+p_a)\\
&=0,
\end{align*}
where the second equality derives from (\ref{eq:user1_margin}).
Thus, we have the following optimal bonus:
\begin{equation}\label{eq:optimal_bonus}
B_a^\mathbbm{2}=q_a-p_a.
\end{equation}
Note that under bonus in (\ref{eq:optimal_bonus}), the optimal revenue is
\begin{equation}\label{eq:optimal_revenue}
R_{a}^\mathbbm{2}=B_a^\mathbbm{2} t_a^\mathbbm{2}.
\end{equation}
One can check  that when $B_a=q_a-p_a$ and $t_a$ given by (\ref{eq:s_min_one}), owner $1$ will not come to platform $a$,
because,
\begin{align*}
&U_{1a}-U_{1b}\\
=&[p_a s_{1a}+B_a(s_{1a}-t_a)]-[p_b s_{1a}+\int_{1-s_{1a}}^{1-s_{1b}}(f'_1(x)-p_b)\ud x]\\
=&[(p_a+B_a-p_b)s_{1a}-\int_{1-s_{1a}}^{1-s_{1b}}(f'_1(x)-p_b)\ud x]\\
&-[(p_a+B_a-p_b)s_{2a}-\int_{1-s_{2a}}^{1-s_{2b}}(f'_2(x)-p_b)\ud x]\\
<&0.
\end{align*}
That is, the optimal bonus to attract owner $2$ does not attract owner $1$.
In this case, platform $a$ gets a positive revenue of $R_{a}^\mathbbm{2}$.

\textbf{Step (iv): Derive the optimal strategy for platform $a$ to maximize revenue, given that it attracts both owner $1$ and owner $2$}.  The revenue for platform $a$ will then be:
\begin{align*}
R_{a}^{both}=&(q_a-p_a) s_{1a}-B_a(s_{1a}-t_a)\\
&+(q_a-p_a) s_{2a}-B_a(s_{2a}-t_a)\\
\leq&(q_a-p_a-B_a)s_{2a}+(3p_a+B_a-2p_b)s_{1a}\\
&-2\int_{1-s_{1a}}^{1-s_{1b}}(f_1'(x)-p_b)\ud x.
\end{align*}

Note that $R_{a}^{both}=(q_a-p_a-B_a)s_{1a}+(q_a-p_a-B_a) s_{2a} + 2B_a t_a>0$ as long as $q_a-p_a\geq B_a\geq p_b-p_a$.
Similarly, the revenue achieves the upper bound when $t_a^{both}=\frac{1}{B_a}[(p_a+B_a-p_b)s_{1a}-\int_{1-s_{1a}}^{1-s_{1b}}(f_1'(x)-p_b)\ud x]$, i.e.,
\begin{align*}
R_{a}^{both}(B_a)=&(q_a-p_a-B_a)s_{2a}+(q_a+p_a+B_a)s_{1a}\\
&-2p_bs_{1b}-2\int_{1-s_{1a}}^{1-s_{1b}}f_1'(x)\ud x.
\end{align*}
Taking the derivative of $R_{a}^{both}(B_a)$  with respect to $B_a$ and setting it to zero, we have:
\begin{align*}
\frac{dR_{a}^{both}}{dB_a}
=& (q_a-p_a-B_a)(\frac{ds_{2a}}{dB_a}+\frac{ds_{1a}}{dB_{1}})+s_{1a}-s_{2a}\\
=&0.
\end{align*}
Thus,
\begin{align}\label{eq:b*}
B^*=p_a-\frac{s_{2a}-s_{1a}}{\frac{ds_{2a}}{dB_a}+\frac{ds_{1a}}{dB_{1}}}<p_a.
\end{align}
This gives the optimal bonus to attract both owners:
\begin{equation}\label{eq:b_both}
B_a^{both}=
\begin{cases}
B^* &   p_a+B^*>p_b\\
p_b-p_a &   \text{otherwise},
\end{cases}
\end{equation}
and the corresponding minimum service required for subsidies:
\begin{equation}
t_a^{both}=
\begin{cases}
\frac{1}{B^*}[(p_a+B^*-p_b)s_{1a}\\
-\int_{1-s_{1a}}^{1-s_{1b}}(f_1'(x)-p_b)\ud x] &  p_a+B^*>p_b\\
0 & \text{otherwise}.
\end{cases}
\end{equation}


In conclusion, When platform $a$ uses a linear loyalty program and platform $b$ does not, platform $a$ can have strict positive revenue. The optimal subsidizing strategy is as follow.
\begin{enumerate}
	\item If $R_{a}^\mathbbm{2}>R_{a}^{both}(B_a^{both})$,
	platform $a$'s equilibrium strategy is $(B_a,t_a)=(p_a,t_a^\mathbbm{2})$, i.e., targeting only owner $2$.
	\item Otherwise, platform $a$ sets loyalty program to be $(B_a,t_a)=(B_{a}^{both},t_a^{both})$, i.e., wining both owners.
\end{enumerate}

\subsection{Proof of Theorem \ref{thm:vs}}
We now analyze the case when $p=p_a=p_b$ and the platforms compete using different loyalty programs, i.e.,  one platform adopts the linear loyalty program and the other introduces sign-up bonus.
Let platform $a$ use linear loyalty program with per-unit sharing reward $B_a$ and minimum sharing threshold $t_a$, and let platform $b$ adopt a sign-up bonus $B_b$.

Note that in this case, for all owner $i$, we also have
\begin{align*}
f_i'(x_{ia})&=p+B_a,\\
f_i'(x_{ib})&=p.
\end{align*}

Our proof proceeds in three steps.

\textbf{Step (i): Derive the conditions for platform $a$ to have positive revenue}.
If platform $a$ targets only owner $2$, the utilities for owner $2$ on different platforms are given by:
\begin{align*}
U_{2a}&=f_2(x_{2a})+p s_{2a}+B_a(s_{2a}-t_a)\\
U_{2b}&=f_2(x_{2b})+p s_{2b}+B_b.
\end{align*}
We thus need:
\begin{align*}
U_{2a}-U_{2b} =&B_a(s_{2a}-t_a)-B_b-\int_{1-s_{2a}}^{1-s_{2b}}(f_2'(x)-p)dx\\
\geq & 0,
\end{align*}
which gives an upper bound for the minimum service $t^{\mathbbm{2}}_{a}$:
\begin{equation}
B_a t^{\mathbbm{2}}_{a}\leq B_a s_{2a}-B_b-\int_{1-s_{2a}}^{1-s_{2b}}(f_2'(x)-p)dx. \label{eq:both_side_bs}
\end{equation}

Similarly,  if  platform $a$ wants to attract owner $1$,
we have the following tighter condition of $t_a$, which will imply that owner $2$ also comes to the platform, i.e.,
\begin{equation}\label{eq:one_side_bs}
B_a t^{both}_{a}\leq B_a s_{1a}-B_b -\int_{1-s_{1a}}^{1-s_{1b}}(f_1'(x)-p)dx.
\end{equation}

\textbf{Step (ii): Derive the conditions for  platform $b$ to have positive revenue.}
The first part is to derive a lower bound of $B_b$ for platform $b$ to have owner $1$.
First note that the utilities for owner $1$ at each platform are given by:
\begin{align*}
U_{1a}&=f_1(x_{1a})+B_a(s_{1a}-t_a)+p s_{1a}\\
U_{1b}&=f_2(x_{1b})+B_b+p s_{1b}.
\end{align*}
We want:
\begin{align*}
&U_{1b}-U_{1a}\\
=&f_1(x_{1b})+B_b+p s_{1b}-f_1(x_{1a})-B_a(s_{1a}-t_a)-p s_{1a}\\
=&\int_{1-s_{1a}}^{1-s_{1b}}(f_1'(x)-p)dx+B_b-B_a(s_{1a}-t_a)\\
\geq & 0,
\end{align*}
which gives a lower bound for $B_b^{\mathbbm{1}}$:
\begin{equation}
B_b^{\mathbbm{1}}\geq B_a(s_{1a}-t_a)-\int_{1-s_{1a}}^{1-s_{1b}}(f_1'(x)-p)dx. \label{eq:b2_bound}
\end{equation}
Similarly, the lower bound of $B_b$ for platform $b$ to have owner $2$ will be higher, and if bonus $B_b$ satisfies the following lower bound, platform $b$ will have both owners.
\begin{equation}\label{eq:b2_bound12}
B_b^{both} \geq B_a(s_{2a}-t_a)-\int_{1-s_{2a}}^{1-s_{2b}}(f_2'(x)-p)dx.
\end{equation}

From the above, we can now show that platform $b$ must have $B_b\leq \frac{q-p}{2}(s_{1b}+s_{2b})$ in order to maintain positive revenue.
The reason is that if $B_b> \frac{q-p}{2}(s_{1b}+s_{2b})$ and  platform $b$ has both owner $1$ and $2$, the revenue becomes
\begin{align*}
R&=(q-p)(s_{1b}+s_{2b})-2B_b\\
&< (q-p)(s_{1b}+s_{2b})-2\cdot \frac{q-p}{2}(s_{1b}+s_{2b})\\
&=0.
\end{align*}
If instead platform $b$ only has owner $1$, the revenue will be:
\begin{align*}
R&=(q-p) s_{1b}-B_b\\
&<(q-p) s_{1b}-\frac{q-p}{2}(s_{1b}+s_{2b})\\
&<0.
\end{align*}

\textbf{Step (iii) Derive the condition for platform $a$ to ``squeeze out''  platform $b$.}
Consider platform $a$'s strategy to be $B_a=q-p$, and always have $t_a\leq t^{\mathbbm{2}}_{a}$, which guarantees $(s_{2a}-t_a)^+=s_{2a}-t_a$.
In this case, let $\psi$ denote platform $b$'s revenue difference  between  having both owners and only owner $1$.
\begin{align*}
\psi=&R_{b}^{both}-R_{b}^{\mathbbm{1}}\\
=&(q-p)s_{2b}-2\bigg(B_a (s_{2a}-t_{a})-\int_{1-s_{2a}}^{1-s_{2b}}(f_2'(x)-p)\ud x\bigg)\\
&+\bigg(B_a (s_{1a}-t_{a})^+-\int_{1-s_{1a}}^{1-s_{1b}}(f_1'(x)-p)\ud x\bigg).
\end{align*}
That is, if $\psi\geq 0$, platform $b$ will choose a higher sign-up bonus to win both owners. Otherwise, it will only try to attract owner $1$.

If $\psi\geq 0$, let platform $a$ choose $t=t^{\mathbbm{2}}_{a}$  according to (\ref{eq:both_side_bs}), by assuming the worst case that $B_b=\frac{q-p}{2}(s_{1b}+s_{2b})$:
\begin{equation}
B_a t_a = B_a s_{2a}-\frac{q-p}{2}(s_{1b}+s_{2b})-\int_{1-s_{2a}}^{1-s_{2b}}(f_2'(x)-p)dx. \label{eq:p1_sm}
\end{equation}
Note that the right hand side above is always positive.
Thus, platform $a$ will have a positive revenue $B_a t_a$.
In this case, if platform $b$ wants to attract both owners, it  needs  $B_b > \frac{q-p}{2}(s_{1b}+s_{2b})$, which will result in a negative revenue $R_{b}^{both}<0$.

On the other hand, if platform $b$ only wants to attract owner $1$,  it  can lower its bonus such that (\ref{eq:b2_bound}) is satisfied. Using  (\ref{eq:p1_sm}), the revenue of platform $b$ is given by:
\begin{align*}
R_{b}^{\mathbbm{1}}=R_{b}^{both} -\psi \leq 0.
\end{align*}
Hence, if $\psi\geq 0$, platform $b$ will always have no revenue given platform $a$'s strategy in (\ref{eq:p1_sm}).


If $\psi <0$, platform $a$ can choose threshold $t'_{a}=t^{both}_{a}$ according to (\ref{eq:one_side_bs}) by assuming that $B_b=(q-p)s_{1b}$ to tighten the constraint such that:
\begin{equation}
B_a t'_{a} = B_a s_{1a}-(q-p) s_{1b}-\int_{1-s_{1a}}^{1-s_{1b}}(f_1'(x)-p)dx. \label{eq:p1_sm_2}
\end{equation}
Note that the right hand side  is always positive. Thus, platform $a$ will have a positive revenue of $B_a t'_{a}$.

In this case, if platform $b$ only  attracts owner $1$, it needs $B_b \geq (q-p) s_{1b}$, which leads to negative revenue $R_{b}^{\mathbbm{1}}=(q-p) s_{1b}-B_b \leq 0$.
If instead platform $b$ attracts both owners $1$ and $2$, it needs $B_b$ to satisfy (\ref{eq:b2_bound12}), and the revenue becomes:
\begin{align*}
R_{b}^{both}
=&R_{b}^{\mathbbm{1}}+\psi <0.
\end{align*}
The implication is that when $\psi<0$, platform $b$ still has zero revenue given platform $a$'s strategy in (\ref{eq:p1_sm_2}).

\subsection{Proof of Theorem \ref{thm:exclusive} }
Suppose there exists some owner $i$ with strategy $(x_i,s_{ia},s_{ib})$ such that $s_{ia},s_{ib}>0$. If $s_{ia}\geq t_a$, we show that sharing only on platform $a$ is better, i.e.,
\begin{align*}
&U_i(x_i,s_{ia}+s_{ib},0)-U_i(x_i,s_{ia},s_{ib})\\
=&[f_i(x_i)+p_a (s_{ia}+s_{ib})+B_a (s_{ia}+s_{ib}-t_a)]\\
&-[f_i(x_i)+p_a s_{ia}+p_b s_{ib}+B_a (s_{ia}-t_a)]\\
=& (p_a+B_a-p_b)s_{ib}\\
>&0.
\end{align*}
Similarly, if $s_{ia}< t_a$, we must have that sharing only on platform $b$ is better, i.e.,
\begin{align*}
&U_i(x_i,0,s_{ia}+s_{ib})-U_i(x_i,s_{ia},s_{ib})\\
=&[f_i(x_i)+p_b (s_{ia}+s_{ib})]-[f_i(x_i)+p_a s_{ia}+p_b s_{ib}]\\
=& (p_b-p_a)s_{ia}\\
>&0.
\end{align*}
Thus, owner $i$ will never choose actions such that $s_{ia},s_{ib}>0$.


\end{document}